# Spin-orbit coupling induced splitting of Yu-Shiba-Rusinov states in antiferromagnetic dimers

Philip Beck[1], Lucas Schneider[1], Levente Rózsa[2†], Krisztián Palotás[3,4,5], András Lászlóffy[3,5], László Szunyogh[5,6], Jens Wiebe[1*] and Roland Wiesendanger[1]

[1] Department of Physics, University of Hamburg, D-20355 Hamburg, Germany

[2] Department of Physics, University of Konstanz, D-78457 Konstanz, Germany

[3] Institute for Solid State Physics and Optics, Wigner Research Center for Physics, H-1525 Budapest, Hungary

[4] MTA-SZTE Reaction Kinetics and Surface Chemistry Research Group, University of Szeged, H-6720 Szeged, Hungary

[5] Department of Theoretical Physics, Budapest University of Technology and Economics, H-1111 Budapest, Hungary

[6] MTA-BME Condensed Matter Research Group, Budapest University of Technology and Economics, H-1111 Budapest, Hungary

Contact information: *jwiebe@physnet.uni-hamburg.de, †levente.rozsa@uni-konstanz.de

## Abstract

Magnetic atoms coupled to the Cooper pairs of a superconductor induce Yu-Shiba-Rusinov states (in short Shiba states). In the presence of sufficiently strong spin-orbit coupling, the bands formed by hybridization of the Shiba states in ensembles of such atoms can support low-dimensional topological superconductivity with Majorana bound states localized on the ensembles' edges. Yet, the role of spin-orbit coupling for the hybridization of Shiba states in dimers of magnetic atoms, the building blocks for such systems, is largely unexplored. Here, we reveal the evolution of hybridized multi-orbital Shiba states from a single Mn adatom to artificially constructed ferromagnetically and antiferromagnetically coupled Mn dimers placed on a Nb(110) surface. Upon dimer formation, the atomic Shiba orbitals split for both types of magnetic alignment. Our theoretical calculations attribute the unexpected splitting in antiferromagnetic dimers to spin-orbit coupling and broken inversion symmetry at the surface. Our observations point out the relevance of previously unconsidered factors on the formation of Shiba bands and their topological classification.

# Main

The interplay between spin-orbit coupling (SOC), magnetism and superconductivity has been extensively studied in recent years due to their applications in quantum computation, particularly concerning the realization of topological qubits based on Majorana bound states (MBS). Evidence of MBS that can exist on the edges of topological superconductors have been reported in various systems involving strong SOC, ranging from semiconductor nanowires proximity coupled to *s*-wave superconductors[1–4], over magnetic vortex cores in topological superconductors[5,6], to one[7–11]- and two-dimensional[12–14] magnetic nanostructures on s-wave superconductors. Promising building blocks for the latter systems are states formed by the hybridization of Yu-Shiba-Rusinov excitations (referred to as Shiba states)[15–17] which lead to the emergence of so-called Shiba bands in nanostructures. Shiba states are induced in the vicinity of magnetic impurities embedded in or adsorbed on the surface of a superconductor via a potential that locally breaks Cooper pairs. Aiming at tailoring the Shiba bands for topological superconductivity, experimental work has focused on investigations of the Shiba states of single magnetic impurities on superconducting substrates[18–26] and of coupled dimers of such impurities[27–30].

In dimers with spacings less than the lateral extent of the Shiba states, the bound states are expected to hybridize. As calculated in ref.[17], there is a fundamental difference between ferromagnetically (FM) and antiferromagnetically (AFM) aligned dimers. In FM dimers, the states strongly hybridize and split into a symmetric and an antisymmetric linear combination of the single-impurity Shiba states. In contrast, for AFM alignment, the hybridization is expected to be weaker since quasiparticles of opposite spin are scattered preferentially by the two impurities, which leads to a smaller shift in the Shiba state energies. Importantly, the two Shiba states remain degenerate in a perfectly AFM-aligned dimer, since exchanging the positions of the two impurities while simultaneously switching the spin directions is a symmetry of the system[31–33]. Experimental results have partially confirmed this picture by observing the presence and the absence of the splitting in dimers which have been identified as FM- and AFM-aligned in density-functional theory calculations, respectively[28]. Accordingly, in the absence of information about the exchange interaction between the localized spins[27,30], it was argued that the observation of the splitting of Shiba states is sufficient to exclude an AFM coupling. All of these experimental observations have been explained based on the theoretical framework formulated by Yu, Shiba and Rusinov[15–17]. This theory does not take into account SOC, and its influence on the Shiba states has been considered in surprisingly few works so far[34,35]. Over the recent decades, a plethora of novel phenomena in solid-state physics has been demonstrated to arise due to the combination of SOC with inversion-symmetry breaking. These include the emergence of Rashba-split surface states in the electronic structure[36]; the mechanism of the Dzyaloshinsky–Moriya interaction[37,38], which gives rise to chiral non-collinear magnetic configurations[39–41], the formation of MBS in magnetic chains proximity coupled to a superconductor[42]; and the presence of the crystal anomalous Hall effect in collinear antiferromagnets[43].

Here, we reveal a so far unconsidered mechanism of Shiba state hybridization caused by SOC in noncentrosymmetric systems. We present a scanning tunneling spectroscopy (STS) study of the multi-orbital Shiba states of single Mn adatoms and Mn dimers on Nb(110). Using the tip of a scanning tunneling microscope (STM) to artificially construct dimers, we vary interatomic orientations and spacings. We identify dimers both with FM and AFM alignments based on spin-polarized measurements. Regardless of the relative orientation of the spins, we observe shifted and split Shiba states in the Mn dimers. However, for the AFM case, the spatial distributions of their wavefunctions no longer clearly resemble the usual symmetric and antisymmetric combinations of the single-impurity Shiba states that are found for the FM case. Our theoretical calculations demonstrate that taking into account SOC and inversion-symmetry breaking is necessary for lifting the twofold degeneracy of Shiba states in AFM oriented dimers. We argue that considering this phenomenon is essential for understanding the subgap excitations in artificially designed magnetic nanostructures at the surfaces of superconductors.

**Multi-orbital Shiba states of single Mn adatoms**

Mn atoms were deposited on a clean Nb(110) surface (Fig. 1c, see Methods) and are adsorbed on the hollow site in the center of four Nb atoms (Fig. 1e). First, we revisit the d$I$/d$V$ spectra of the single adatoms with a considerably better energy resolution than reported previously[18] (Fig. 1a,b) which is achieved using superconducting tips and lower temperatures (see Methods). Compared to the spectra taken on bare Nb(110), they reveal four pairs of additional resonances inside the superconductor's energy gap, one at positive and one at negative bias symmetrically with respect to the Fermi energy $E_F$ ($V$=0 V), which we label ±α, ±β, ±γ and ±δ. The resonance labeled ±β is only visible as a shoulder of the ±α peak. Using d$I$/d$V$ maps, we determine the spatial distribution[19,20,30] of these four resonances, revealing an astonishing resemblance to the shape of the well-known atomic $d$-orbitals (Figs. 1f-i, see the corresponding maps of the positive bias partner in Supplementary Fig. 4). The energetically highest and most intense state ±α has a circular shape and faint lobes along the [001] (*x*) and [1$\bar{1}$0] (*y*) direction (Fig. 1f and Supplementary Fig. 4a), which hints towards an origin in the $d_{z^2}$ or the $d_{x^2-y^2}$ orbital. The three other states have $d_{xy}$-like (±β), $d_{xz}$-like (±γ) and $d_{yz}$-like (±δ) shapes (Figs. 1g-i and Supplementary Figs. 4b-d). Moving away from the adatom center, the spectral intensities of the states decrease rapidly to a tenth of their maximum values within a range of 1 nm, and only very weak oscillations of the spectral intensity can be observed at larger distances (Supplementary Figs. 4e-l). We correspondingly assign the states to multi-orbital Shiba states [19,20,30] formed by the Mn adatom as can be understood based on the following theoretical model. The free-standing Mn atom has a spin of $S = \frac{5}{2}$ in the ground state according to Hund's first rule, with each of its five degenerate *d*-orbitals being singly occupied. As the atom is placed on the Nb(110) surface, its atomic states hybridize with the substrate, and the degeneracy of the states is lifted by the crystal field. Based on symmetry arguments (see Supplementary Note 9 for details), it can be concluded that the $d_{xy}$, $d_{xz}$ and $d_{yz}$ orbitals function as three separate scattering channels of different shapes and strengths acting on the quasiparticles of the superconductors. The $d_{z^2}$ and $d_{x^2-y^2}$ orbitals hybridize and form two scattering channels, only one of which leads to an observable Shiba state in the experiments, while the other may be hidden in the coherence peaks. The shapes of these scattering channels were extracted from *ab initio* calculations performed for the Mn adatom on Nb(110), based on the procedure described in ref.[20] (see Methods and Supplementary Note 5). The strengths of the non-magnetic ($K$) and magnetic ($JS/2$) scattering were determined in such a way that the calculated local density of states (LDOS) at the position of the adatom, presented in Fig. 1j, resembles the experimental d*I*/d*V* spectrum, including the energy positions of the peaks and the particle-hole asymmetry in intensity between Shiba pairs located at positive and negative bias, respectively. The values of $K$ and $JS/2$ are listed in Supplementary Table 1. The spatial distributions of the LDOS at the Shiba resonance energies, illustrated in Figs. 1k-n, reasonably agree with the experimental data (Figs. 1f-i) demonstrating the robustness of the theoretical model.

**Hybridization of Shiba states in ferromagnetic dimers**

We now turn to the investigation of the Shiba states in Mn dimers. We can tune the magnetic exchange interaction between FM and AFM by laterally manipulating one of the adatoms with the STM tip (see Methods), thereby varying the crystallographic direction and interatomic spacing. We first consider the close-packed dimer along the $[1\bar{1}0]$ direction (see Fig. 2c,d) denoted as $\sqrt{2}a - [1\bar{1}0]$ dimer, where $a = 329.4$ pm is the nearest-neighbor spacing along [001], corresponding to the bulk lattice constant of bcc Nb. Spin-polarized STM (SP-STM) measurements on close-packed chains in this direction (Supplementary Note 3 and Supplementary Fig. 3) indicate that the spins in this dimer are FM-aligned (Fig. 2d). The d$I$/d$V$ spectrum taken on top of the dimer (Figs. 2a and b, blue lines) shows six pairs of Shiba states in contrast to the four pairs of states observed for the single adatom (black line). By comparing their energies and spatial distributions in d$I$/d$V$ maps (Figs. 2e-j and Supplementary Fig. 5) with the energies and shapes of the single-adatom Shiba states (Figs. 1b,f-i), we conjecture that these 6 states can be sorted out into 3 pairs upon hybridization of single-adatom Shiba states, one of the $\pm\alpha$ (Figs. 2e,f and Supplementary Figs. 5a,b), one of the $\pm\gamma$ (Figs. 2g,h and Supplementary Figs. 5c,d), and one of the $\pm\delta$ (Figs. 2i,j and Supplementary Figs. 5e,f) state. Moreover, for each of these 3 pairs, one state has maxima in the *xz* plane in the center between the two Mn adatoms of the dimer (Figs. 2e,g,I and Supplementary Figs. 5a,d,e) while the other approximately has a nodal line in that plane (Figs. 2f,h,j and Supplementary Figs. 5b,c,f). Therefore, we tentatively assign them to symmetric (s) and antisymmetric (a) linear combinations of the single-adatom $\pm\alpha$, $\pm\gamma$, and $\pm\delta$ Shiba states[17,31,32]. Hybridized states of the type $\pm\beta$ could not be identified in the dimer, presumably because of their weak intensity as observed for the single adatom.

Model calculations (see Methods) using the scattering channel parameters determined for the adatom support these conclusions (Fig. 2k-s). There are eight pairs of Shiba states of the dimer visible as peaks in the calculated LDOS (Fig. 2k), which may be separated into symmetric and antisymmetric combinations with respect to the *xz* mirror plane, as it was performed for the experimental images. Based on the spatial profiles of the states (Figs. 2l-s) we denote them as $\pm\alpha_s$ and $\pm\alpha_a$ (Figs. 2n,o), $\pm\gamma_s$ and $\pm\gamma_a$ (Figs. 2p,q), as well as $\pm\delta_s$ and $\pm\delta_a$ (Figs. 2r,s), respectively. The two additional states (Figs. 2l,m) which are not observed in the experiment are assigned to the $\pm\beta_s$ and $\pm\beta_a$ states, although their spatial profile also shows similarities with the $\alpha$ states being close by in energy. The latter states were separated from each other by performing calculations with a higher energy resolution than shown in Fig. 2k. Comparing experimental and theoretical results of the energetic shifts of the hybridized Shiba states relative to the single-adatom states and the splitting of symmetric and antisymmetric states (Tab. 1), we can conclude that the model reproduces the experimental results reasonably well.

## Hybridization of Shiba states in antiferromagnetic dimers

To investigate the effect of the spin configuration on Shiba states of a dimer and to check for the reported absence of split Shiba states in AFM-coupled dimers[31,32], we study the nearest-neighbor dimer constructed along the diagonal of the centered rectangular unit cell (denoted as $\sqrt{3}a/2 - [1\bar{1}1]$ dimer, Figs. 3b,c). Recent SP-STM measurements on short chains of close-packed Mn adatoms along this direction[44] indicate that the exchange interaction between the adatoms in this dimer is AFM (Fig. 3c). Surprisingly, the d$I$/d$V$ spectrum taken on the dimer (Fig. 3a, blue line) displays six pairs of sharp peaks, implying the hybridization and energetic splitting of the single-adatom Shiba states. From the spatial distributions of the d$I$/d$V$ maps taken at the peak positions (Figs. 3d-i and Supplementary Fig. 6) and by following similar arguments as for the $\sqrt{2}a - [1\bar{1}0]$ dimer given above, we conclude that also for this $\sqrt{3}a/2 - [1\bar{1}1]$ dimer the ±α, ±γ, and ±δ Shiba states of the single Mn adatom split into pairs of hybridized states. However, while the pairs ±$\alpha_1$ and ±$\alpha_2$ (Figs. 3d,e and Supplementary Figs. 6a,b) as well as ±$\delta_1$ and ±$\delta_2$ (Figs. 3h,i and Supplementary Figs. 6e,f) still resemble symmetric and antisymmetric linear combinations of the single-adatom Shiba states, the situation is not as obvious for the pair ±$\gamma_1$ and ±$\gamma_2$ (Figs. 3f,g and Supplementary Figs. 6c,d). Note, that the symmetry in this dimer is reduced from $C_{2v}$ to $C_2$ containing only a twofold rotation around the $z$-axis, which slightly complicates the assignment of the states. Similar to the $\sqrt{2}a - [1\bar{1}0]$ dimer, spectroscopic signatures of the possible ±β Shiba orbitals in this dimer could not be identified, presumably because of their reduced intensity. Remarkably, the observed splitting of Shiba states is at least of similar strength for the AFM- as for the FM-coupled dimer (cf. Tabs. 1 and 2). A Shiba state splitting of similar strength is as well observed for the $2a - [001]$ dimer which also is AFM-coupled (Supplementary Fig. 7). For the $2a - [001]$ dimer the split Shiba states likewise no longer simply resemble symmetric and antisymmetric linear combinations of the single-adatom Shiba states (Supplementary Note 4). In contrast, each Shiba state at positive energy appears rather antisymmetric compared to their negative-energy partners which are rather symmetric. Overall, it is surprising to find Shiba state splitting in AFM dimers since theoretical calculations in the seminal paper of Rusinov[17], extended upon in later works[31,32,45], predicted that for this type of coupling the Shiba states may shift in energy, but they always remain degenerate.

**Role of spin-orbit coupling for the hybridization of Shiba states**

In order to find a theoretical explanation for this experimental observation, we first discuss the origin of the degeneracy of the Shiba states in AFM-coupled dimers based on symmetry arguments. In the absence of magnetic impurities, the system may be characterized by a Hamiltonian which is invariant under time reversal, represented for spin-1/2 systems as the antiunitary operator $T = i\sigma_4^y \tau^z K$, where $K$ denotes complex conjugation and $\sigma_4^y \tau^z$ is a spin matrix in Nambu space (see Methods). According to Kramers' theorem $T^2 = -1$ implies that all eigenstates of this Hamiltonian are pairwise degenerate. The inclusion of the AFM-aligned impurities breaks time-reversal symmetry, but the Hamiltonian remains invariant under the operation $T_r = TC_2$, a combination of time reversal and a rotation around the out-of-plane direction by 180° in real space. $T_r$ effectively also acts as time reversal, since it is an antiunitary symmetry with $T_r^2 = -1$. This means, that Kramers' theorem still applies in this case, implying a pairwise degeneracy of the Shiba states in the superconductor in particular. This description holds for any AFM-aligned Mn dimer located in the hollow positions on the Nb(110) surface, since they all have a $C_2$ rotational symmetry if the magnetic configuration is disregarded. The situation is different if SOC is taken into account. Time reversal ($T$) remains a symmetry of the system in the absence of magnetic impurities, but $T_r$ must be extended since the bulk Hamiltonian is no longer invariant under spin rotation. The spin and orbital degrees of freedom have to be rotated simultaneously, which results in $T_{r'} = TC_2 i\sigma_4^z \tau^z$ as an antiunitary symmetry operation for an AFM dimer with out-of-plane spins, which is the case for the $\sqrt{3}a/2 - [1\bar{1}1]$ dimer. For this symmetry one obtains $T_{r'}^2 = 1$, meaning that Kramers' theorem does not apply, and a lifting of the degeneracies is expected. Results of a numerical calculation using material-specific parameters for the AFM $\sqrt{3}a/2 - [1\bar{1}1]$ dimer are displayed in Fig. 4. In the absence of SOC, the LDOS shows the same number of peaks as in the case of the adatom (Fig. 4a), although their energies are shifted. The spatial distributions of these dimer Shiba states are visualized in Figs. 4b-e, where they were assigned the same ±α, ±β, ±γ and ±δ labels based on their relative similarity in energy position and spatial distribution to the Shiba states of the adatom. As expected from the considerations above, if SOC is taken into account (see Methods), the resonance peaks clearly split in the AFM dimer (Fig. 4f). Since the spatial distributions of the Shiba states in Figs. 4g-n still show remarkable similarity to the states obtained without SOC (cf. Figs. 4b with g, c with i or j, d with k, and e with m or n) we can assign them to the original states and accordingly name them ±α₁, ±α₂, etc. The calculated spatial distributions and splittings including SOC qualitatively agree with the experimental data (cf. Figs. 3d-i and 4g-n and Tab. 2). In particular, as found in the experiment (Figs. 3f,g and Supplementary Figs. 7c-h) the spatial distributions no longer resemble symmetric and antisymmetric linear combinations of the original single-adatom states. Finally, it is worth mentioning that SOC in the presence of inversion-symmetry breaking also gives rise to the Dzyaloshinsky–Moriya interaction, due to which the spin orientation of the FM- and AFM-coupled dimers might become non-collinear. Based on first-principles calculations using the Korringa–Kohn–Rostoker method[46], it was found for the $\sqrt{3}a/2 - [1\bar{1}1]$ dimer that this canting from the collinear AFM state is only around 0.5° (see Supplementary Table 3). This degree of non-collinearity would not be sufficiently large to explain the experimentally observed magnitude of the splittings of the peaks without considering SOC directly in the calculation of the Shiba states (Supplementary Note 8).

**Discussion**

In conclusion, we demonstrated that the Shiba states of a Mn adatom on the Nb(110) substrate hybridize and split in dimers with considerable overlap between the states. This can be observed not only in FM-coupled but also in AFM-coupled dimers, with a similar magnitude of the energy splitting for both cases. Our theoretical analysis attributes this phenomenon to the breaking of an effective time-reversal symmetry of the AFM dimer, which would otherwise protect the degeneracy of the Shiba states, by SOC in the non-centrosymmetric system. Note that this splitting is not expected to occur for impurities in centrosymmetric bulk systems. There, the atoms in the dimer may be exchanged by spatial inversion $P$, rather than only by the $C_2$ rotation discussed above. Since the spins remain invariant under spatial inversion, the $TP$ symmetry would then be sufficient to enforce the degeneracy of the Shiba states. An effective time-reversal symmetry $T_r$ is commonly found not only in antiferromagnetic dimers, but in antiferromagnetic chains and two-dimensional structures as well. The breaking of this symmetry by the SOC in non-centrosymmetric systems should clearly distinguish the emergent Majorana bound states in Shiba bands from their counterparts in time-reversal invariant systems[42,47]. Most importantly, our findings indicate that observing the presence or the absence of the splitting of Shiba states in dimers on surfaces cannot be used as a fingerprint for the type of exchange interaction between two magnetic impurities[28,31]. These results should motivate to revisit previous experimental observations of Shiba states and theoretical predictions on the formation of Majorana bound states at the ends of Shiba atom chains by taking into account the SOC in systems with broken inversion symmetry, shedding new light on these potential building blocks of topological superconductors.

## Methods

**STM and STS measurements**

All experiments were performed in a home-built ultra-high vacuum STM setup, operated at a temperature of 320 mK[48]. STM images were obtained by stabilizing the STM tip at a given bias voltage $V_{bias}$ applied to the sample and tunneling current $I$. d$I$/d$V$ spectra were obtained using a standard lock-in technique with a modulation frequency of $f_{mod}$=4142 Hz, a modulation voltage of $V_{mod}$=20 µV added to $V_{bias}$, and stabilization of the tip at $V_{stab}$=6 mV and $I_{stab}$=1 nA before opening the feedback and sweeping the bias. To visualize spatial distributions of the Shiba states, we defined spatial grids on the structure of interest with a certain pixel resolution, typically with a spacing of about 50 pm to 100 pm between grid points. d$I$/d$V$ spectra were then measured on every point of the grid, with the parameters listed above except for a different modulation voltage of $V_{mod}$=40 µV. d$I$/d$V$ maps are slices of this d$I$/d$V$ grid evaluated at a given bias voltage.

We used an electrochemically etched and in-situ flashed tungsten tip, which was indented a few nanometers into the niobium surface, covering it with niobium and producing a superconducting apex of the tip.

We achieved a considerably better energy resolution as compared to previous results obtained on this sample system[18] by performing experiments at a lower temperature (320 mK) and making use of superconducting probe tips. Based on a d$I$/d$V$ spectrum taken on a patch of clean Nb(110) we find the two substrate coherence peaks at ±2.93 mV as indicated by the red vertical lines in Fig. 1a. From measurements with a normal metal tip, we deduce a superconducting gap of $\Delta_{Nb} = 1.50$ mV (see Supplementary Note 2). Thereby, the shift in the position of the coherence peaks in the measurement with the superconducting STM probe tip can be used to determine its superconducting gap of $\Delta_{tip} = 1.43$ mV[30]. All Shiba states in the d$I$/d$V$ spectra and maps shown here are correspondingly shifted in energy by $\Delta_{tip}$.

**Sample preparation**

The Nb(110) single crystal with a purity of 99.999% was introduced into the ultra-high vacuum chamber and subsequently cleaned by Ar ion sputtering and flashing to about 2400°C to remove surface-near oxygen[49]. Mn atoms were evaporated to the sample while keeping the substrate temperature below 10 K, to achieve a statistical distribution of single adatoms (see Fig. 1c, Supplementary Note 1 and Supplementary Fig. 1a). Combining the knowledge of the position of single Mn adatoms with respect to the Nb(110) surface unit cell visible in atomically resolved STM images (Supplementary Fig. 1b) with atom-manipulation images (Supplementary Fig. 1c) with the fact that we find identical features in d$I$/d$V$ spectra for all single Mn adatoms, as well as with the results of first-principles calculations (Supplementary Note 5), we conclude that the only energetically stable adsorption site for Mn adatoms on Nb(110) is the hollow site in the center of four Nb atoms (Fig. 1e). Artificial dimers were constructed using STM tip-induced atom manipulation with a typical tunneling resistance of about 60 k$\Omega$.

**Model calculations**

The system was described by the Hamiltonian

$$H = H_\mathrm{b} + H_\mathrm{R} + H_\mathrm{sc} + H_\mathrm{imp}, \tag{1}$$

with the spectrum of bulk Nb

$$H_\mathrm{b} = \sum_\mathbf{k} \Phi_\mathbf{k}^\dagger \xi_\mathbf{k} \tau^z \Phi_\mathbf{k}, \tag{2}$$

nearest-neighbor Rashba-type SOC on the bcc(110) surface of Nb

$$H_\mathrm{R} = \sum_\mathbf{k} \Phi_\mathbf{k}^\dagger \left[ 4t_\mathrm{R} \sin\left(\frac{k^x}{2}a\right) \cos\left(\frac{\sqrt{2}}{2}k^y a\right) \sigma_4^y - 4t_\mathrm{R} \sin\left(\frac{\sqrt{2}}{2}k^y a\right) \cos\left(\frac{k^x}{2}a\right) \sqrt{2}\sigma_4^x \right] \Phi_\mathbf{k}, \tag{3}$$

s-wave superconducting pairing

$$H_\mathrm{sc} = \sum_\mathbf{k} \Phi_\mathbf{k}^\dagger \Delta \tau^x \Phi_\mathbf{k}, \tag{4}$$

and the impurity potential

$$H_\mathrm{imp} = \frac{1}{N} \sum_{\mu,\alpha,\mathbf{k},\mathbf{k}'} \Phi_\mathbf{k}^\dagger \Psi_\mu(\mathbf{k}) \left( K^\mu \tau^z - \frac{J^\mu S}{2} S^\alpha \sigma_4^\alpha \tau^z \right) \Psi_\mu^*(\mathbf{k}') \Phi_\mathbf{k}. \tag{5}$$

In Eqs. (1)-(5), $\Phi_\mathbf{k} = 1/\sqrt{2}(c_{\mathbf{k}\uparrow}, c_{\mathbf{k}\downarrow}, c_{-\mathbf{k}\downarrow}^\dagger, c_{-\mathbf{k}\uparrow}^\dagger)$ is the Nambu spinor, with the electron annihilation operators $c_{\mathbf{k}\sigma}$. The matrices in Nambu space read

$$\sigma_4^\alpha = \begin{bmatrix} \sigma^\alpha & 0 \\ 0 & -\sigma^\alpha \end{bmatrix}, \tau^z = \begin{bmatrix} \sigma^0 & 0 \\ 0 & -\sigma^0 \end{bmatrix}, \tau^x = \begin{bmatrix} 0 & \sigma^0 \\ \sigma^0 & 0 \end{bmatrix} \tag{6}$$

where $\sigma^\alpha$ are the usual $2 \times 2$ Pauli matrices for $\alpha=x,y,z$ and $\sigma^0$ is the unit matrix. The bulk Nb spectrum $\xi_\mathbf{k}$ was determined from a Slater–Koster two-center tight-binding model, using the parameters in ref.[50]. The original spectrum containing 9 bands (5$s$, 5$p$ and 4$d$) was mapped to a single-band model only containing states in the energy range [$E_F$-10$\Delta$, $E_F$+10$\Delta$]. This was performed to decrease the numerical cost of the wave vector summations since the contribution of the states far away from the Fermi level to the Shiba states is expected to be small. For a different implementation of the integration in the vicinity of the Fermi surface with an energy cut-off, see ref.[51]. In the Rashba term, $x$ and $y$ are along the [001] and [1$\bar{1}$0] directions, respectively (Fig. 1e). The Rashba parameter $t_R$=7.5 meV was approximated by identifying Rashba-split surface states in Nb(110) from *ab initio* calculations based on the screened Korringa–Kohn–Rostoker (SKKR) method (Supplementary Note 6). The order parameter $\Delta$=1.5 meV was based on the experimentally determined gap size $\Delta_\mathrm{Nb}$ and assumed to be real-valued and homogeneous[17]. In the impurity potential, $S$ is the spin quantum number, $(S^x, S^y, S^z)$ is the direction of the localized magnetic moment, and $\mu$ indexes the different scattering channels, including different orbitals in a single impurity and multiple sites in the case of dimers. $\Psi_\mu(\mathbf{k}) = \mathrm{diag}\left(\psi_\mu(\mathbf{k}), \psi_\mu(\mathbf{k}), \psi_\mu^*(-\mathbf{k}), \psi_\mu^*(-\mathbf{k})\right)$ describes the shapes of the scattering centers, with the wave functions $\psi_\mu$ expressed in a plane-wave basis. The scattering wave functions were identified with the $d$ orbitals of a Mn adatom on Nb(110) calculated using the Vienna Ab-initio Simulation Package (VASP)[52–54], based on the procedure described in ref.[20] (see Supplementary Note 5 for details). The $K^\mu$ and $J^\mu S/2$ parameters were determined by approximating the experimentally observed energy positions and the asymmetries between negative and positive bias for all Shiba states in the adatom. The same values were used for calculating the Shiba states in the dimers. The magnetic and non-magnetic scattering parameters with SOC (used in Figs. 1j-n, 2k-s and 4f-n) and without it (used in Figs. 4a-e) are listed in Supplementary Tables 1 and 2, respectively. An approximation of the scattering parameters based on the density of states determined from SKKR calculations is discussed in Supplementary Note 7.

The LDOS was calculated using a Green's function-based method[28,51,55]. The Green's function at complex energy $z$ is expressed as

$$G(z, \mathbf{k}, \mathbf{k}') = G_0(z, \mathbf{k})\delta_{\mathbf{k},\mathbf{k}'} + G_0(z, \mathbf{k}) T(z, \mathbf{k}, \mathbf{k}') G_0(z, \mathbf{k}) \tag{7}$$

where

$$G_0(z,\mathbf{k}) = \left(z - \xi_{\mathbf{k}}\tau^z - 4t_R \sin\left(\frac{k^x}{2}a\right)\cos\left(\frac{\sqrt{2}}{2}k^y a\right)\sigma_4^y + 4t_R \sin\left(\frac{\sqrt{2}}{2}k^y a\right)\cos\left(\frac{k^x}{2}a\right)\sqrt{2}\sigma_4^x - \Delta\tau^x\right)^{-1} \quad (8)$$

is the Green's function of the substrate and

$$T(z,\mathbf{k},\mathbf{k}') =$$
$$\sum_{\mu,\mu'} \Psi_\mu(\mathbf{k})\frac{1}{N}\left(K^\mu\tau^z - \frac{J^\mu S}{2}S^\alpha\sigma_4^\alpha\tau^z\right)\left[I_4 - \sum_{\mathbf{k}''}\Psi_\mu^*(\mathbf{k}'')G_0(z,\mathbf{k}'')\Psi_{\mu'}(\mathbf{k}'')\frac{1}{N}\left(K^{\mu'}\tau^z - \frac{J^{\mu'}S}{2}S^\alpha\sigma_4^\alpha\tau^z\right)\right]^{-1}\Psi_{\mu'}^*(\mathbf{k}') \quad (9)$$

is the $T$ operator describing scattering of the impurities. $I_4$ is the unit matrix in Nambu space. Shiba states are found at the real energies $|z| < \Delta$ inside the gap where $T$ has poles, meaning that the determinant of the matrix in the brackets equals 0. Finding the zeroes of the determinant made it possible to separate the Shiba states very close in energy in Figs. 2k and 4f, and to confirm the degeneracies for the AFM dimer in the absence of spin–orbit coupling in Fig. 4a.

The LDOS was calculated as

$$\text{LDOS}(E,\mathbf{r}) = -\frac{1}{\pi}\text{Im Tr}\left[\frac{1}{N}\sum_{\mathbf{k},\mathbf{k}'} e^{-i(\mathbf{k}-\mathbf{k}')\mathbf{r}} G(E+i\delta,\mathbf{k},\mathbf{k}')\frac{I_4 + \tau^z}{2}\right] \quad (10)$$

after Fourier transformation back into real space and projecting the spectral function on the particle part of Nambu space. In Eq. (10), δ denotes a small imaginary part, which was set to $k_B \cdot 300$ mK to approximate the experimentally observed energy resolution. The spectra in Figs. 1j, 2k, 4a and 4f were obtained by calculating a weighted average of the LDOS over different positions,

$$\text{LDOS}_{\text{spectra}}(E) = \sum_{i=1}^{9} e^{-\frac{r_i}{a}}\text{LDOS}(E,\mathbf{r}_i), \quad (11)$$

with $\mathbf{r}_i = a\left(\cos\frac{i\pi}{4}, \sin\frac{i\pi}{4}, 0\right)$ for $i = 1,\dots,8$ and $\mathbf{r}_9 = (0,0,0)$, measured with respect to the position of an impurity. The averaging was performed since some of the orbitals have a very small spectral weight when measured directly on the adatom due to their symmetry.

## Data availability

The authors declare that the data supporting the findings of this study are available within the paper and its supplementary information files.

## Code availability

The analysis codes that support the findings of the study are available from the corresponding authors on reasonable request.

## Acknowledgements

P.B., R.W., and J.W. gratefully acknowledge support by the SFB 925 'Light induced dynamics and control of correlated quantum systems' of the Deutsche Forschungsgemeinschaft (DFG). L.S., R.W., and J.W. gratefully acknowledge funding by the Cluster of Excellence 'Advanced Imaging of Matter' (EXC 2056 - project ID 390715994) of the Deutsche Forschungsgemeinschaft (DFG). R.W. gratefully acknowledges funding of the European Union via the ERC Advanced Grant ADMIRE (grant No. 786020). L.R. gratefully acknowledges funding from the Alexander von Humboldt Foundation. Financial supports of the National Research, Development, and Innovation Office of Hungary (NKFIH) under Projects No. FK124100 and No. K131938, and of the BME Nanotechnology and Materials Science TKP2020 IE grant of NKFIH Hungary (BME IE-NAT TKP2020) are gratefully acknowledged by A.L., K.P., L.R. and L.Sz.. We acknowledge fruitful discussions with Thore Posske, Stephan Rachel and Dirk Morr.

## Author contributions

P.B., L.S. and J.W. conceived the experiments. P.B. and L.S. performed the measurements and analyzed the experimental data together with J.W.. P.B. and L.R. prepared the figures. K.P. performed the VASP calculations. A.L. performed the SKKR calculations and discussed the data with K.P., L.R. and L.Sz.. L.R. performed the model calculations. P.B., L.R., and J.W. wrote the manuscript. All authors contributed to the discussions and the corrections of the manuscript.

## Competing interests

The authors declare no competing interests.

# Figures

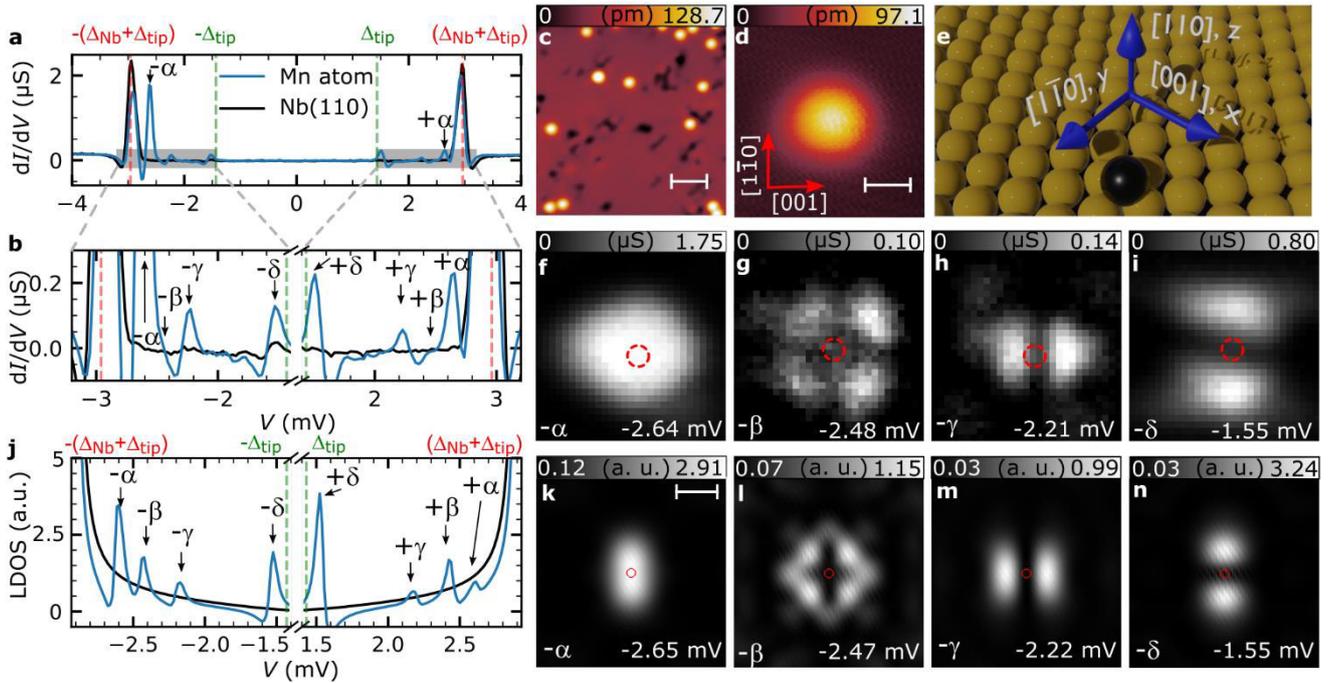

**Figure 1 | Shiba states of single Mn adatoms on Nb(110). a**, d$I$/d$V$ spectra obtained on bare Nb(110) (black) and over a Mn adatom (blue). Red and green vertical lines mark the positions of the coherence peaks ±($\Delta_{Nb}$+$\Delta_{tip}$) and the tip gap $\Delta_{tip}$, respectively. **b**, Magnification of the spectra shown in panel **a**. The coherence peaks and the two peaks with the largest intensity are left out for the sake of visibility. Shiba states are labeled and marked by arrows. **c**, Overview STM image of the Mn/Nb(110) sample, where bright protrusions are single Mn adatoms and black depressions correspond to residual oxygen. The white scale bar has a length of 3 nm. ($V_{bias}$=100 mV and $I$=200 pA). **d**, STM image of a single Mn adatom, which was used for recording the spectra in **a** and **b**, as well as for the d$I$/d$V$ maps in **f-i**. The white scale bar has a length of 500 pm and the red arrows point along two high symmetry directions [1$\bar{1}$0] and [001] of the Nb(110) surface. The directions also apply to panels **f-i** and **k-n**. **e** 3D rendered illustration of the position of the single Mn adatom (black sphere) with respect to the Nb substrate atoms (yellow spheres), including the high-symmetry directions indicated by blue arrows with their nomenclature. **f-i**, d$I$/d$V$ maps measured at the given bias voltages of the Shiba states marked in **a** and **b** and in the area of **d** in the vicinity of the single Mn adatom[56]. **j**, Calculated LDOS inside the superconducting gap convoluted with the DOS of the superconducting tip ($\pm\Delta_{tip}=1.43$ mV) at the position of the Mn adatom (blue) and in the bare superconductor (black). **k-n**, Calculated two-dimensional maps of the LDOS at the bias voltages indicated in the figures. The white scale bar has a length of 500 pm. Red circles in **f-i** and **k-n** highlight the position of the adatom.

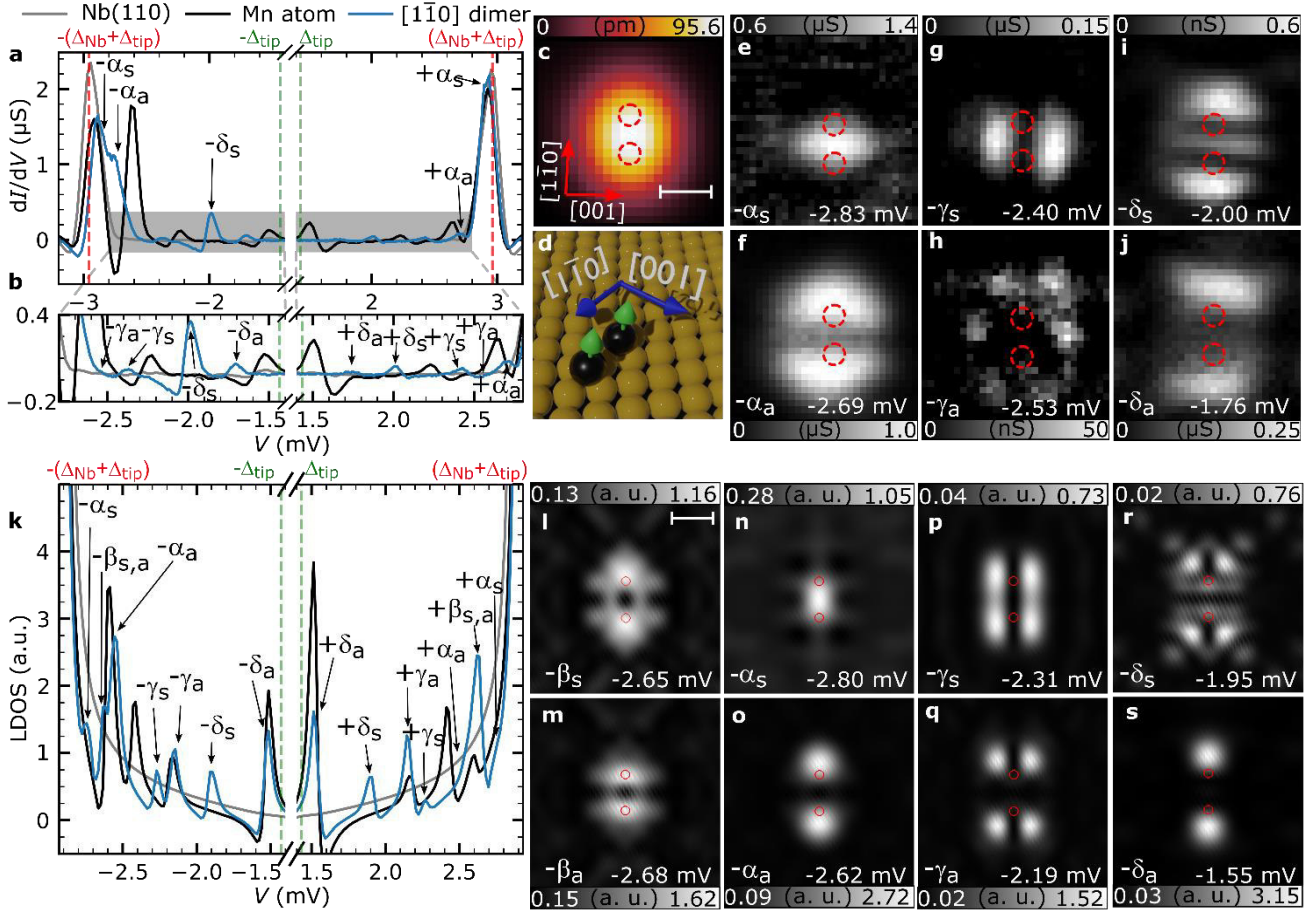

**Figure 2 | Hybridized Shiba states in a FM-coupled $\sqrt{2}a - [1\bar{1}0]$ Mn dimer. a,** d$I$/d$V$ spectrum taken on top of a $\sqrt{2}a - [1\bar{1}0]$ dimer (blue) with reference spectra for the substrate (gray) and a single Mn adatom (black). **b,** Magnification of the spectra shown in panel **a**. Shiba states are labeled and marked by arrows. Vertical lines mark sample and tip gaps. **c,** Small-scale STM image of a $\sqrt{2}a - [1\bar{1}0]$ Mn dimer. The white scale bar has a length of 500 pm ($V_{bias}$=6 mV and $I$=1 nA). The directions and the scale bar also apply to panels **e-j**. **d,** 3D rendered illustration of the positions of the two Mn adatoms (black spheres) in the investigated dimer with respect to the Nb substrate atoms (yellow spheres), including the spin structure indicated by green arrows. **e-j,** d$I$/d$V$ maps measured on the dimer in the same area shown in panel **c** for the given bias voltages where a peak/shoulder is visible in the spectrum in **a** and **b**[56]. **k,** LDOS calculated on one adatom of the FM-aligned $\sqrt{2}a - [1\bar{1}0]$ dimer (blue), on a single adatom (black) and on the bare superconductor (gray) convoluted with the DOS of the superconducting tip. **l-s,** Calculated two-dimensional maps of the LDOS at the indicated bias voltages. The white scale bar has a length of 500 pm; crystallographic directions as in panel **c**. Red circles in **c**, **e-j** and **l-s** denote the locations of the Mn adatoms in the dimer.

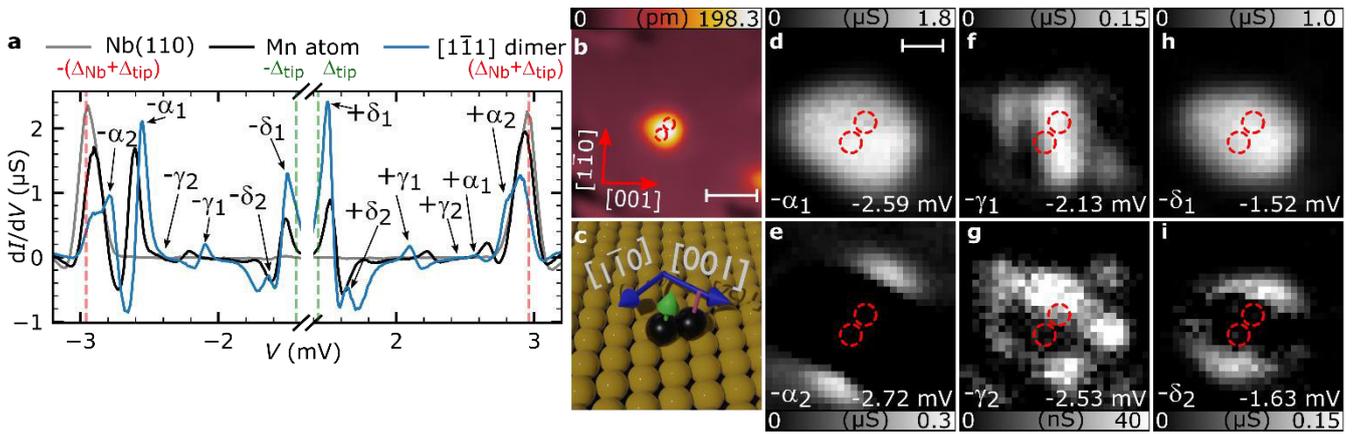

**Figure 3 | STS of hybridized Shiba states in an AFM-coupled $\sqrt{3}a/2 - [1\bar{1}1]$ Mn dimer. a,** d$I$/d$V$ spectrum taken on a $\sqrt{3}a/2 - [1\bar{1}1]$ Mn dimer (blue). A reference spectrum taken on a single Mn adatom (black) and a patch of clean Nb(110) (gray) are pasted to the background. Shiba states are labeled and marked by arrows. **b,** STM image of a $\sqrt{3}a/2 - [1\bar{1}1]$ Mn dimer. The white scale bar has a length of 1 nm ($V_{bias}$=6 mV and $I$=1 nA). **c,** 3D rendered illustration of the positions of the two Mn adatoms (black spheres) in the investigated dimer with respect to the Nb substrate atoms (yellow spheres), including the spin structure indicated by green and red arrows. **d-i,** d$I$/d$V$ maps measured on the dimer shown in panel **b** for the given bias voltages where a peak/shoulder is visible in the spectrum in **a**[56]. The same orientation from panel **b** applies to these maps. The white scale bar has a length of 500 pm. Red circles in **b** and **d**-**i** denote the locations of the Mn adatoms in the dimer.

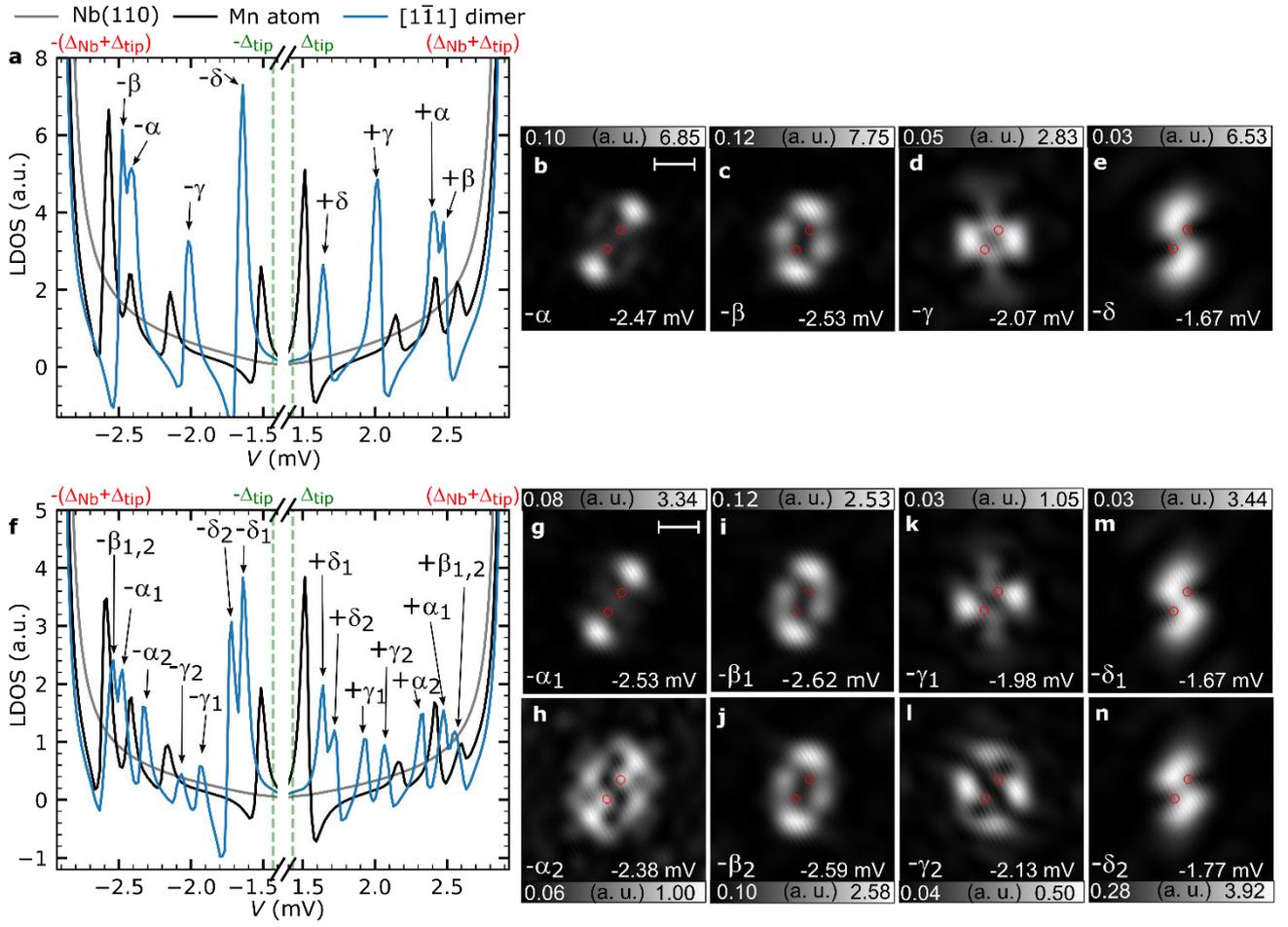

**Figure 4 | Calculation of hybridized Shiba states in an AFM-coupled $\sqrt{3}a/2 - [1\bar{1}1]$ Mn dimer.** LDOS inside the superconducting gap calculated on one adatom in the out-of-plane AFM-aligned dimer (blue), for a single adatom (black) and on the bare superconductor (gray), **a** without and **f** with taking SOC into account. The spectrum is convoluted with the superconducting DOS of the tip. Shiba states are labeled and marked by arrows. Two-dimensional maps of the LDOS at the indicated bias voltages show the spatial profiles of the states without (**b**-**e**) and with (**g**-**n**) SOC. Red circles denote the positions of the adatoms in the dimer. The white scale bar has a length of 500 pm. The crystallographic axes are the same as in Fig. 3b. Magnetic and non-magnetic scattering parameters with and without SOC are given in Supplementary Tables 1 and 2, respectively.

# Tables

| Shiba state | Experiment | | Theory | |
|---|---|---|---|---|
| | Shift (μV) | Splitting (μV) | Shift (μV) | Splitting (μV) |
| α | +120 | +140 | +60 | +180 |
| β | - | - | +195 | -30 |
| γ | +260 | -130 | +30 | +120 |
| δ | +330 | +240 | +200 | +400 |

**Table 1 | Comparison of energetic shifts and splittings of hybridized Shiba states in a FM-coupled $\sqrt{2}a - [1\bar{1}0]$ Mn dimer.** The energetic shifts were calculated from the experimental and theoretical results by $\frac{1}{2}(|E_{n_a}| + |E_{n_s}|) - |E_n|$, where $E_n$ are the energetic positions of the single-adatom Shiba states $n \in (\alpha, \beta, \gamma, \delta)$, and $E_{n_a}$ and $E_{n_s}$ are the energies of the respective antisymmetric and symmetric Shiba states in the dimer. The splittings of hybridized Shiba states are calculated by $|E_{n_s}| - |E_{n_a}|$.

| Shiba state | Experiment | | Theory | |
|---|---|---|---|---|
| | Shift (μV) | Splitting (μV) | Shift (μV) | Splitting (μV) |
| α | +15 | -130 | -195 | +150 |
| β | - | - | +135 | +30 |
| γ | +120 | -400 | -165 | -150 |
| δ | +25 | -110 | +170 | -100 |

**Table 2 | Comparison of energetic shifts and splittings of hybridized Shiba states in an AFM-coupled $\sqrt{3}a/2 - [1\bar{1}1]$ Mn dimer.** The energetic shifts were calculated from the experimental and theoretical results by $\frac{1}{2}(|E_{n_2}| + |E_{n_1}|) - |E_n|$, where $E_n$ are the energetic positions of the single-adatom Shiba states $n \in (\alpha, \beta, \gamma, \delta)$, and $E_{n_2}$ and $E_{n_1}$ are the energies of the respective hybridized Shiba states in the dimer. The splittings of hybridized Shiba states are calculated by $|E_{n_1}| - |E_{n_2}|$.

# Spin-orbit coupling induced splitting of Yu-Shiba-Rusinov states

# in antiferromagnetic dimers

Philip Beck[1], Lucas Schneider[1], Levente Rózsa[2†], Krisztián Palotás[3,4,5], András Lászlóffy[3,5], László Szunyogh[5,6], Jens Wiebe[1*] and Roland Wiesendanger[1]

[1] Department of Physics, University of Hamburg, D-20355 Hamburg, Germany

[2] Department of Physics, University of Konstanz, D-78457 Konstanz, Germany

[3] Institute for Solid State Physics and Optics, Wigner Research Center for Physics, H-1525 Budapest, Hungary

[4] MTA-SZTE Reaction Kinetics and Surface Chemistry Research Group, University of Szeged, H-6720 Szeged, Hungary

[5] Department of Theoretical Physics, Budapest University of Technology and Economics, H-1111 Budapest, Hungary

[6] MTA-BME Condensed Matter Research Group, Budapest University of Technology and Economics, H-1111 Budapest, Hungary

Contact information: *jwiebe@physnet.uni-hamburg.de, †levente.rozsa@uni-konstanz.de

**Supplementary Note 1. Sample preparation**

An overview image of the sample surface, which was used for the experiments, is shown in Supplementary Fig. 1a. Using the procedure described in the Methods section of the main text, we obtained a similar Nb(110) surface quality as previously reported in the literature[1]. Only a few dark depressions, which can be identified as oxygen impurities, remain on the surface. Clean patches of Nb(110), at least 5 nm x 5 nm in size, could easily be found and were used for the scanning tunneling spectroscopy (STS) experiments in order to avoid any influence of the oxygen impurities. Supplementary Fig. 1b shows an atomically resolved STM image. White lines tracing the rows of surface Nb atoms along the two crystallographic directions are added as a guide to the eye. We find that the centers of the Mn adatoms (bright protrusions) lie in the rectangular area defined by two neighboring lines along the $[001]$ and two neighboring lines along the $[1\bar{1}0]$ direction. An atom-manipulation image[2], which is obtained by pushing or dragging a single Mn adatom across the sample surface while recording an STM image at high tunneling conductance, is shown in Supplementary Fig. 1c. We find that the symmetry and the distances of the protrusions in the manipulation image follow that of the four-fold coordinated hollow sites of the Nb(110) surface. In combination with the information drawn from Supplementary Fig. 1b and VASP calculations (Supplementary Note 5), we conclude that this hollow site, for which a Mn adatom adsorbs in the diamond shaped center of four surface Nb atoms, is the only favorable and observable adsorption site.

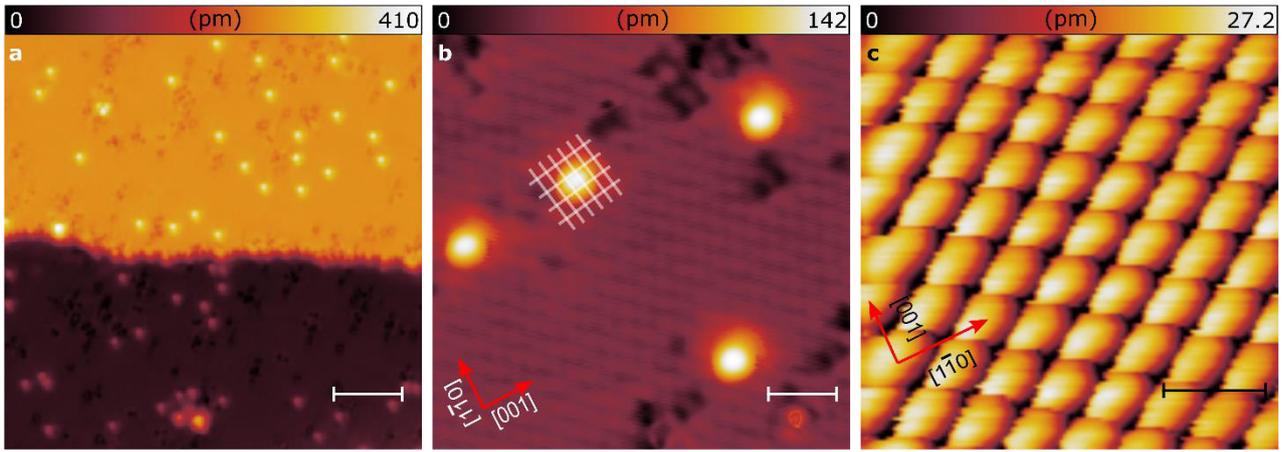

**Supplementary Figure 1 | Overview of Mn/Nb(110) sample and Mn adsorption site. a**, Large-scale STM image providing an overview of the prepared Mn/Nb(110) sample. A small amount of oxygen contaminations remained on the sample, which are visible as dark depressions. The white scale bar represents a length of 5 nm ($V_{bias}$=6 mV and $I$=200 pA). **b**, Atomically resolved STM image of the Nb(110) surface including four single Mn adatoms in the scan frame. White lines trace the rows of surface Nb atoms along the $[001]$- and $[1\bar{1}0]$-directions. The white scale bar represents a length of 2 nm ($V_{bias}$=6 mV and $I$=1 nA). **c**, Atom-manipulation image obtained by dragging/pushing a Mn adatom across the Nb(110) surface while taking the image. The black scale bar represents a length of 500 pm ($V_{bias}$=-3 mV and $I$=50 nA).

## Supplementary Note 2. Determining the tip gap

Accurately knowing the size of the superconducting gap of the STM tip is important if one wants to determine the exact energetic positions of the Shiba states, since all features are shifted by $\pm\Delta_{\text{tip}}$ to higher bias voltages. One way to characterize the superconducting gap of the tip is based on a comparison with STS measurements using a normal conducting metal tip. An example of a tunneling spectrum obtained with a normal conducting Cr tip on a Nb(110) sample is shown in Supplementary Fig. 2a. The spectrum is fitted by standard STS theory using a Dynes function in order to model the substrate's local density of states (LDOS),

$$\rho(E = eV) = \rho_0 \cdot \text{Re}\left(\frac{E-i\gamma}{\sqrt{(E-i\gamma)^2-\Delta^2}}\right). \qquad \text{Supplementary Equation (1)}$$

The resulting superconducting gap of Nb is $\Delta_{Nb}$=1.50 mV, which is indicated by the red vertical lines. Since we now know the actual superconducting gap $\Delta_{Nb}$ of the sample measured at 320 mK, we can calculate the superconducting gap $\Delta_{tip}$ of the tip used in all other measurements. We recall the positions of the coherence peaks indicated in Fig. 1a of the main text (±2.93 mV) which are given by ±($\Delta_{Nb}$+$\Delta_{tip}$). By a simple subtraction of the fitted gap of Nb ($\Delta_{Nb}$) from this quantity, we obtain a tip gap of $\Delta_{tip} = 1.43$ mV. This value is substantiated by a corresponding fit of a substrate spectrum taken with the superconducting tip assuming two Dynes functions (Supplementary Equation 1), one for the substrate's LDOS for which we now know the fitting parameters, which are then used to calculate a convolution matrix (see ref.[3]), and one for the tip's DOS with a different set of parameters, still to be optimized. The multiplication of the convolution matrix with the tip DOS is fitted to the d$I$/d$V$ spectrum measured on bare Nb(110) which is shown in Supplementary Fig. 2b.

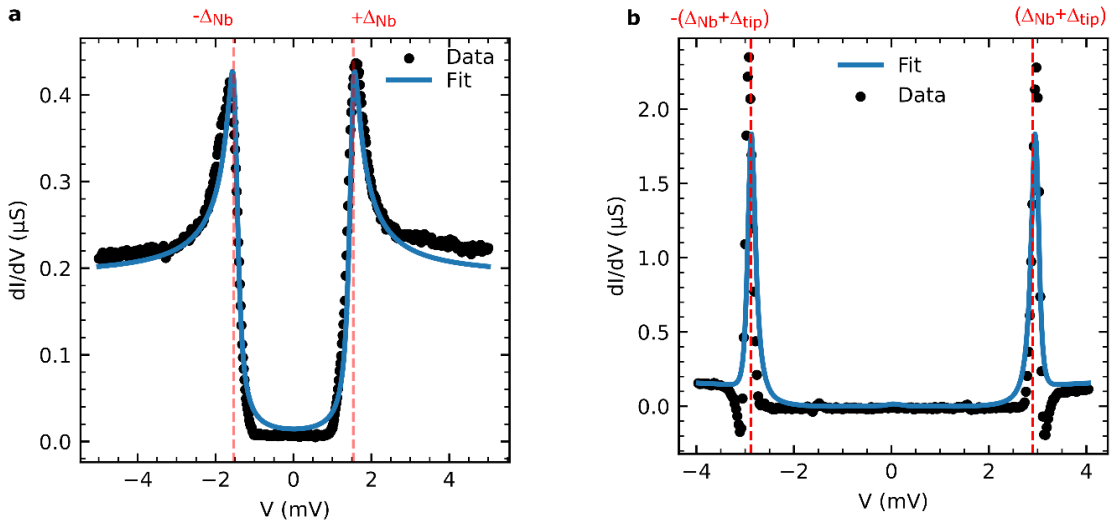

**Supplementary Figure 2 | Determination of the substrate and tip gaps. a,** d$I$/d$V$ spectrum (black dots) taken on bare Nb(110), using a non-superconducting Cr tip. The spectrum was fitted (blue curve) assuming a broadened Dynes function, Supplementary Equation (1), as the LDOS of the superconducting Nb(110). The extracted fitting parameters are $\Delta_{Nb} = 1.50$ mV and $\gamma = 0.12$ mV. Red vertical lines indicate the position of ±$\Delta_{Nb}$ ($V_{stab}$=-5 mV, $I_{stab}$=1 nA, $V_{mod}$=40 µV, Cr tip). **b,** d$I$/d$V$ spectrum (black dots) taken on bare Nb(110), using the superconducting tip, which was used for all measurements discussed in the main text. The spectrum was fitted (blue curve) assuming two broadened Dynes functions, one for the substrate's LDOS with fixed parameters (see **a**) and one for the tip's DOS (Supplementary Equation (1)). The extracted fitting parameters are $\Delta_{tip} = 1.43$ mV and $\gamma = 300$ nV. Red vertical lines indicate the positions of ±($\Delta_{Nb} + \Delta_{tip}$) ($V_{stab}$=6 mV, $I_{stab}$=1 nA, $V_{mod}$=40 µV, Nb-coated W tip).

**Supplementary Note 3. Spin-polarized STM measurements**

A recent publication on spin-polarized STM (SP-STM) of Mn chains assembled atom-by-atom using tip-induced atom manipulation on clean Nb(110) discusses the experimentally observed magnetic coupling for a chain of Mn adatoms spaced by $\sqrt{3}a/2$ and built along the $[1\bar{1}1]$-direction and for a chain of Mn adatoms spaced by 2a built along the [001]-direction[4]. In both of these chains the antiferromagnetic state observed in a small magnetic field indicates that the neighboring Mn adatoms are AFM-coupled.

To demonstrate the FM coupling of Mn adatoms with a spacing of $\sqrt{2}a$ along the $[1\bar{1}0]$-direction, the results of an SP-STM measurement using Cr bulk tips on a chain of 11 Mn adatoms with nearest-neighbor separation of $\sqrt{2}a$ along the $[1\bar{1}0]$-direction are shown in Supplementary Fig. 3. The same microtip and measurement conditions as in ref.[4] were used for this measurement. An STM image of the Mn$_{11}$ chain is shown in Supplementary Fig. 3a. d$I$/d$V$ maps taken on the same area across this chain in +0.5 T and -0.5 T magnetic field, i.e. pointed along opposite directions, are shown in Supplementary Figs. 3b and c, respectively. The contrast of the color scale is adjusted to the same values for both images. One finds that the Nb(110) substrate has the same color in both magnetic fields (compare edges of Supplementary Figs. 3b and c). However, on top of the chain one finds different signals due to the tunneling magnetoresistance. In a magnetic field of +0.5 T, the d$I$/d$V$ signal on the chain is lower than it is in a magnetic field of -0.5 T, indicating that the chain's overall magnetization follows the external magnetic field and is aligned parallel and antiparallel to the stable Cr tip magnetization. Since the d$I$/d$V$ signal is constant along the chain for each magnetic field, we conclude that the chain is in a ferromagnetically ordered state.

The asymmetry in the measured differential conductance obtained from the d$I$/d$V$ maps in opposite magnetic fields is given by:

$$A_{\text{SP}} = \frac{dI/dV(V,\vec{r}_t)|_{-0.5\text{T}} - dI/dV(V,\vec{r}_t)|_{+0.5\text{T}}}{dI/dV(V,\vec{r}_t)|_{-0.5\text{T}} + dI/dV(V,\vec{r}_t)|_{+0.5\text{T}}},$$  Supplementary Equation (2)

where $V$ is the tunneling bias voltage, and $\vec{r}_t$ is the position of the tip apex[5]. A map of this asymmetry calculated from Supplementary Figs. 3b and c is shown in Supplementary Fig. 3d and clearly shows FM order. Therefore, we conclude that Mn adatoms with a spacing of $\sqrt{2}a$ along the $[1\bar{1}0]$-direction are FM coupled.

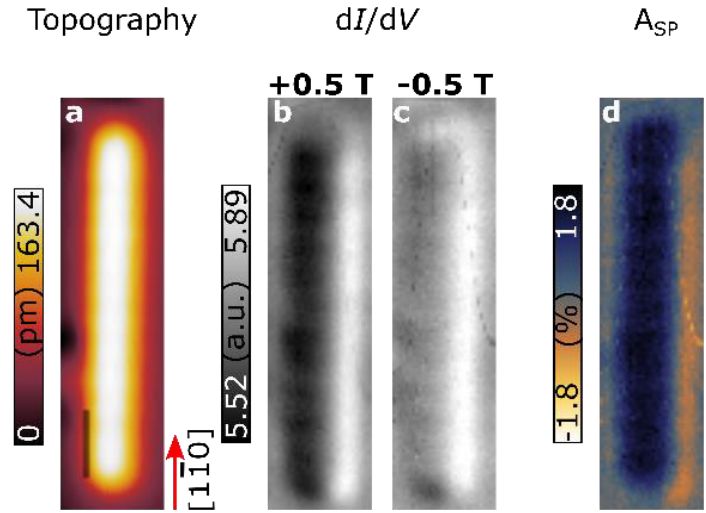

**Supplementary Figure 3 | Magnetic structure of a chain of 11 Mn adatoms with a spacing of $\sqrt{2}a$ along the $[1\bar{1}0]$-direction. a**, STM image of the Mn$_{11}$ chain, which was constructed using atom manipulation. **b** and **c**, are the corresponding d$I$/d$V$ maps measured with a Cr bulk tip at opposite magnetic fields of +0.5 T and -0.5 T ($V_{bias}$=6 mV, $I$=1 nA, $V_{mod}$=2 mV). Both maps are adjusted to the same color scale for comparability. **d**, Asymmetry ($A_{SP}$) map calculated from the d$I$/d$V$ maps in opposite magnetic fields (**b**, **c**) according to Supplementary Equation (2). The black scale bar in **a** represents a length of 1 nm and is valid for all panels.

**Supplementary Note 4. Complementary spectroscopic data**

The following figures, Supplementary Figs. 4a-d, Supplementary Fig. 5 and Supplementary Fig. 6 contain the positive bias counterparts of the d$I$/d$V$ maps shown in the main text of a single adatom (Figs. 1f-i), a $\sqrt{2}a - [1\bar{1}0]$ dimer (Figs. 2e-j) and a $\sqrt{3}a/2 - [1\bar{1}1]$ dimer (Figs. 3d-i), respectively. They allow for the same conclusions regarding the hybridization of Shiba states which were drawn in the main text and support these findings.

Additionally, Supplementary Figs. 4e-l show larger scale (5 nm x 5 nm) d$I$/d$V$ maps measured on a single Mn adatom, to display the strong spatial decay of the spectral intensity of the Shiba states. To make the long-range extension visible at all, the stabilization current was increased to $I_{stab} = 3$ nA, compared to $I_{stab} = 1$ nA which was used for all other d$I$/d$V$ maps. Furthermore, the contrast had to be adjusted.

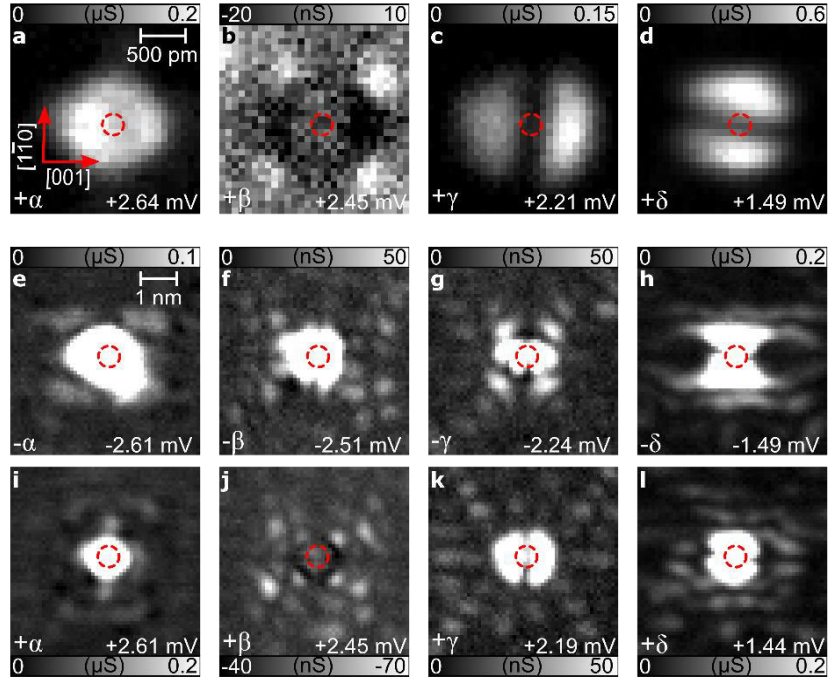

**Supplementary Figure 4 | Positive bias and large-scale d$I$/d$V$ maps of Shiba states of a single Mn adatom.**
**a-d**, Positive bias counterpart of the d$I$/d$V$ maps shown in Figs. 1f-i in the main text of a single Mn adatom taken at the bias voltages as indicated ($V_{stab}$=6 mV, $I_{stab}$=1 nA, $V_{mod}$=40 µV). Red circles denote the position of the adatom. **e-l**, Large-scale d$I$/d$V$ maps (5 nm × 5 nm), recorded with bias voltages as indicated and obtained with a lower stabilization height compared to all other d$I$/d$V$ maps shown in this work. The color maps are adjusted to lower d$I$/d$V$ values in order to enhance the visibility of the long-range extension of the Shiba states ($V_{stab}$=6 mV, $I_{stab}$=3 nA, $V_{mod}$=40 µV).

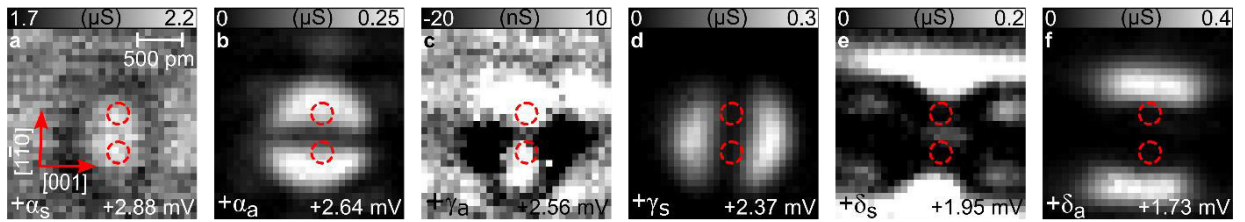

**Supplementary Figure 5 | Hybridization of Shiba states in the FM-coupled $\sqrt{2}a - [1\bar{1}0]$ Mn dimer.**
**a-f**, Positive bias counterpart of the d$I$/d$V$ maps shown in Figs. 2e-j of the main text for the $\sqrt{2}a - [1\bar{1}0]$ Mn dimer at the respective bias voltages, where Shiba states were identified ($V_{stab}$=6 mV, $I_{stab}$=1 nA, $V_{mod}$=40 µV). Red circles denote the positions of the Mn adatoms in the dimer.

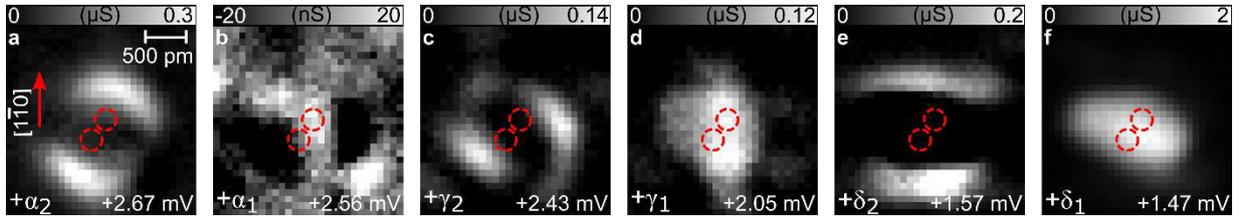

**Supplementary Figure 6 | Hybridization of Shiba states in the AFM-coupled $\sqrt{3}a/2 - [1\bar{1}1]$ Mn dimer. a-f**, Positive bias counterpart of the d$I$/d$V$ maps shown in Figs. 3d-i of the main text for the $\sqrt{3}a/2 - [1\bar{1}1]$ Mn dimer at the respective bias voltages, where Shiba states were identified ($V_{stab}$=6 mV, $I_{stab}$=1 nA, $V_{mod}$=40 µV). Red circles denote the positions of the Mn adatoms in the dimer.

In Supplementary Fig. 7 we present another example of split Shiba states in an AFM-coupled dimer, i.e., the $2a - [001]$ Mn dimer. A d$I$/d$V$ spectrum taken on a single Mn adatom and one taken on the $2a - [001]$ dimer is shown in Supplementary Fig. 7a. A small-scale STM image of the dimer is shown in Supplementary Fig. 7b. d$I$/d$V$ maps, taken at all bias voltages where a peak can be identified in Supplementary Fig. 7a are shown in Supplementary Figs. 7c-h. The spatial distribution of the Shiba state at about ±2.50 mV (Supplementary Figs. 7c and f) consists of two lobes facing away from the dimer center along the [001] direction and reveals a clear nodal line between the two adatoms. In combination with the spatial distribution of the Shiba state at a bias voltage of -2.08 mV (Supplementary Fig. 7g), which has its maximum in intensity in the center between the two Mn adatoms of the dimer, we can conclude that these two pairs of Shiba states have their origin in the split ±$\alpha$ state, and correspondingly name them ±$\alpha_2$ and ±$\alpha_1$. Note that the spatial distribution of the positive bias state at +2.16 mV (Supplementary Fig. 7d), differs from its negative bias counterpart, i.e., it again has a node between the two adatoms of the dimer.

Finally, the spatial distributions of the Shiba states at about ±1.42 mV (Supplementary Figs. 7e and h) resemble that of two ±$\delta$ states of two Mn adatoms, which are not split into hybridized combinations. The Shiba states of AFM-coupled dimers cannot be described as symmetric and antisymmetric combinations of the atomic orbitals, and it may occur that the same state has an intensity maximum in the middle at negative bias and an intensity minimum at positive bias, as can be seen in Supplementary Figs. 7d and g, or 7e and h. Another possibility is that there are two states close in energy, a rather symmetric combination which is more dominant at negative bias and a rather antisymmetric one which is more dominant at positive bias.

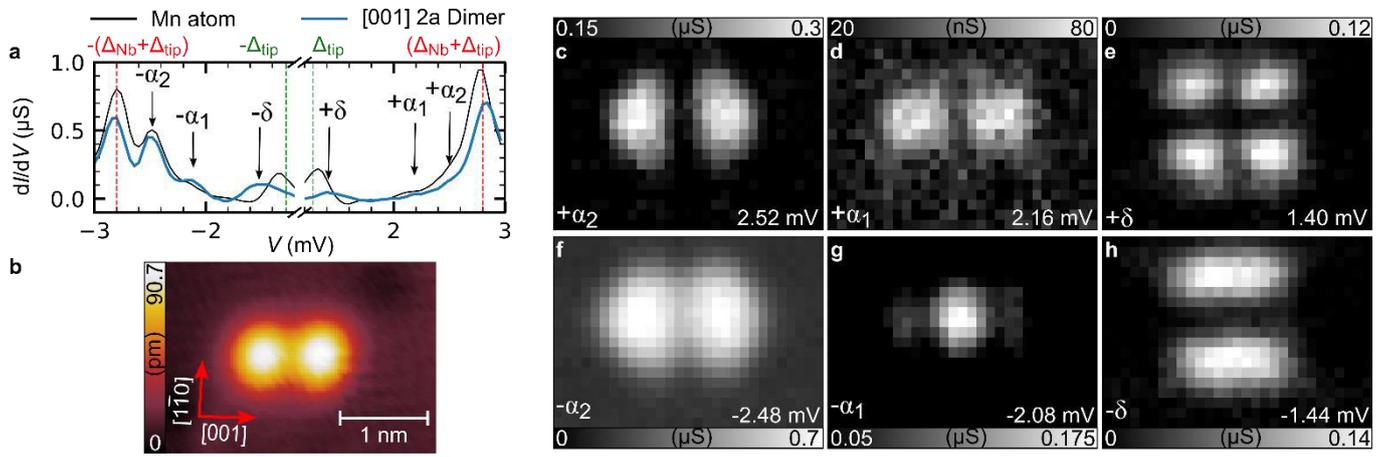

**Supplementary Figure 7 | STS study of hybridized Shiba states of an AFM-coupled 2a − [001] Mn dimer. a**, d$I$/d$V$ spectra obtained in the middle of the two adatoms forming the 2a − [001] Mn dimer (blue) and a reference spectrum taken on a single adatom (black). The superconducting energy gap of the tip is smaller in these spectra ($\Delta_{\text{tip}}$ = 1.275 mV) compared to all other single spectra and d$I$/d$V$ maps displayed in this work. **b**, STM image of the 2a − [001] dimer ($V_{\text{bias}}$=-3 mV and $I$=1 nA). The red arrows indicate the crystallographic directions of the Nb(110) surface. **c-h,** d$I$/d$V$ maps displaying the spatial distributions of the six Shiba states marked in **a**. The lattice vectors and the lateral scale bar marked in **b** are valid for these maps as well. ($V_{\text{stab}}$=-6 mV, $I_{\text{stab}}$=1 nA, $V_{\text{mod}}$=40 µV).

## Supplementary Note 5. VASP calculations

The equilibrium geometry of the Mn adatom on Nb(110) and the shapes of the scattering potentials were determined using the Vienna Ab-initio Simulation Package (VASP)[6–8]. The considered supercell included a single Mn adatom deposited in a hollow position on 4 layers of Nb, with $7 \times 7$ atoms in each layer in the bcc(110) geometry, using the bulk lattice constant $a = 330.04$ pm. A sufficiently thick vacuum region of close to 10 Å was considered in the direction perpendicular to the surface to minimize interactions between repeated supercells. Exchange–correlation effects were included using the potential construction by Perdew, Burke and Ernzerhof[9], and reciprocal-space calculations were restricted to the Γ point due to the large size of the supercell. The Mn adatom and the atoms in the top Nb layer were allowed to relax vertically. It was found that the average vertical distance between the atoms in the two highest Nb layers decreased to 227.49 pm from the bulk value of $\frac{\sqrt{2}}{2}a = 233.37$ pm, and that the Mn adatom relaxed even further, to a vertical distance of 198.68 pm measured from its nearest Nb neighbours. The spin magnetic moment of the Mn adatom was found to be 3.60 μ$_B$. It was confirmed that the total energy in the hollow adsorption site was about 580 meV lower than on top of the Nb atoms, and that the adatom relaxed back into the hollow position when starting the relaxation with a slightly horizontally displaced position. This confirms the hollow position as the only energetically favorable adsorption site for the Mn adatom on the surface.

The scattering wave functions $\psi_\mu$ indirectly entering Eq. (5) of the main text were extracted from the calculations similar to the procedure discussed in ref.[10]. Using the self-consistently optimized geometry, VASP calculations were performed for a minimal cluster consisting of the Mn adatom and its four nearest-neighbor Nb atoms, sufficient for reproducing the crystal-field splitting of the orbitals in the $C_{2v}$ point group. The distance between the Mn adatoms in repeated supercells was at least 20 Å, sufficiently large to avoid interactions between the periodic images, and a $3 \times 3 \times 1$ **k** mesh was used.

By inspecting the orbital-resolved weights of the wave functions projected on the Mn adatom, it was possible to identify the five occupied 3d states in the majority spin channel, as expected for Mn with a half-filled *d* band. In the minimal cluster, these states are still strongly localized on the Mn adatom. There was a single state for each of the $d_{xy}$, $d_{yz}$ and $d_{xz}$ orbitals, and two of them were a linear combination of the hybridized $d_{z^2}$ and $d_{x^2-y^2}$ states, as expected based on symmetry arguments (see Supplementary Note 9).

The wave function coefficients for the identified states were extracted using the WaveTrans program[11]. The reciprocal space coefficients were transformed to real space[12] in a volume centered on the adatom and being similar to the VASP supercell in size. Then, the $\psi_\mu(\mathbf{k})$ coefficients were determined on the finer grid in the vicinity of the Fermi level where the spectrum $\xi_\mathbf{k}$ was calculated, following the energy cut-off discussed in the Methods section of the main text. Due to the fine grid in reciprocal space, the artificial periodicity of the wave function introduced by the supercell approach is suppressed. The wave vectors were restricted to the first Brillouin zone, avoiding periodicity in reciprocal space, thereby obtaining a smooth wave function in real space rather than peaks localized at the real-space lattice sites. The latter property allows for a finer spatial resolution of the Shiba states compared to what can be achieved in lattice tight-binding models, which was necessary because the experimentally observed extension of the Shiba states is restricted to a few lattice constants.

The non-magnetic $K^\mu$ and magnetic $J^\mu S/2$ scattering parameters were determined from a comparison of the calculated LDOS at the adatom with the experimental spectra, as mentioned in the main text. Four different Shiba states were identified experimentally, each of them belonging to a different irreducible representation of the $C_{2v}$ point group. Therefore, we set the scattering parameters for the wave function with the strongest $d_{x^2-y^2}$ character to zero, so that it does not produce a Shiba state. The $K^\mu$ and $J^\mu S/2$ parameters for the remaining four scattering channels are listed in Supplementary Tables 1 and 2 with and without spin–orbit coupling, respectively. Note that turning the spin–orbit coupling on or off influences the density of states at the Fermi level, thereby requiring a different set of parameters to describe the same energy positions of the Shiba states.

| state | $K^\mu$ (eV) | $J^\mu S/2$ (eV) |
|---|---|---|
| $d_{z^2} = \alpha$ | 0.0360 | 0.0610 |
| $d_{xy} = \beta$ | -0.0010 | 0.1070 |
| $d_{xz} = \gamma$ | 0.0220 | 0.0980 |
| $d_{yz} = \delta$ | -0.0090 | 0.0450 |

**Supplementary Table 1 | Scattering parameters for the different scattering channels with spin–orbit coupling.** These values were used for the calculations shown in Figs. 1j-n, 2k-s and 4f-n of the main text.

| state | $K^\mu$ (eV) | $J^\mu S/2$ (eV) |
|---|---|---|
| $d_{z^2} = \alpha$ | 0.0180 | 0.0330 |
| $d_{xy} = \beta$ | 0.0040 | 0.0780 |
| $d_{xz} = \gamma$ | 0.0300 | 0.0750 |
| $d_{yz} = \delta$ | -0.0060 | 0.0335 |

**Supplementary Table 2 | Scattering parameters for the different scattering channels without spin–orbit coupling.** These values were used for the calculations shown in Figs. 4a-e of the main text.

**Supplementary Note 6. Determining the Rashba parameter based on SKKR calculations**

Using the optimized geometry of the surface layer and the adsorbed adatom obtained from VASP calculations (Supplementary Note 5), the surface electronic structure and the magnetic properties of the Mn adatoms and dimers were investigated in more detail based on the fully relativistic screened Korringa–Kohn–Rostoker (SKKR) method[13]. After performing self-consistent calculations for bulk Nb, an interface consisting of 8 Nb layers and 4 layers of empty spheres along the (110) direction was treated. The self-consistent calculations were performed with 16 energy points on a semicircle contour, using 253 $\mathbf{k}$ points in the irreducible part of the surface Brillouin zone and an angular momentum cutoff of $l_{\max} = 3$.

The Rashba parameter in Eq. (3) of the main text was determined by calculating the Bloch spectral function in the surface Nb layer, defined as[14,15]

$$A_L^B(E, \mathbf{k}_\parallel) = -\frac{1}{\pi} \mathrm{Im} \mathrm{Tr} \int G_{LL}^+(E, \mathbf{k}_\parallel, \mathbf{r}) \mathrm{d}^3 \mathbf{r}, \qquad \text{Supplementary Equation (3)}$$

where $E$ is the energy, $\mathbf{k}_\parallel$ is the in-plane wave vector, $L$ is the layer index, and the trace is taken in angular momentum space. $A_L^B(E, \mathbf{k}_\parallel)$ corresponds to the density of states resolved in energy and in wave vector, where integration of the retarded Green's function $G_{LL}^+$ is only carried out in real space over the atomic sphere.

The Bloch spectral function along the high-symmetry lines in the surface Brillouin zone is displayed in Supplementary Fig. 8. The hexagonal Brillouin zone of the bcc(110) surface is shown in a sketch in the figure. Sharp peaks in the Bloch spectral function in the energy-wave vector plane localized in the few top layers of the Nb(110) surface correspond to surface states, while bulk states appear as a smooth background since the translational symmetry perpendicular to the surface is broken, effectively corresponding to an integration along the $k^z$ component of the wave vector. Surface states may be observed in regions in energy and wave vector space where there are no states in the bulk, for example in the region highlighted by the black ellipse along the $\overline{\Gamma P}$ line (see also ref.[1]). The spin–orbit coupling causes a clearly identifiable Rashba splitting in these surface states. It was confirmed that these resonances disappear in the lower Nb layers, and the splitting is no longer observable if spin–orbit coupling is turned off in the calculations. The Rashba parameter was extracted by calculating the energy splitting of the peaks $\Delta E$ in the surface states, and comparing this value to the expression derived from Eq. (3) in the main text,

$$\Delta E = 2\sqrt{\left[4t_\text{R}\sin\left(\frac{k^x}{2}a\right)\cos\left(\frac{\sqrt{2}}{2}k^y a\right)\right]^2 + \left[4\sqrt{2}t_\text{R}\sin\left(\frac{\sqrt{2}}{2}k^y a\right)\cos\left(\frac{k^x}{2}a\right)\right]^2} \quad \text{Supplementary Equation (4)}$$

It was found that $t_\text{R}$=7.5 meV provides a reasonable agreement with the *ab initio* data in the whole highlighted range, and this value was used in turn for the Shiba state calculations in Figs. 1, 2 and 4 in the main text.

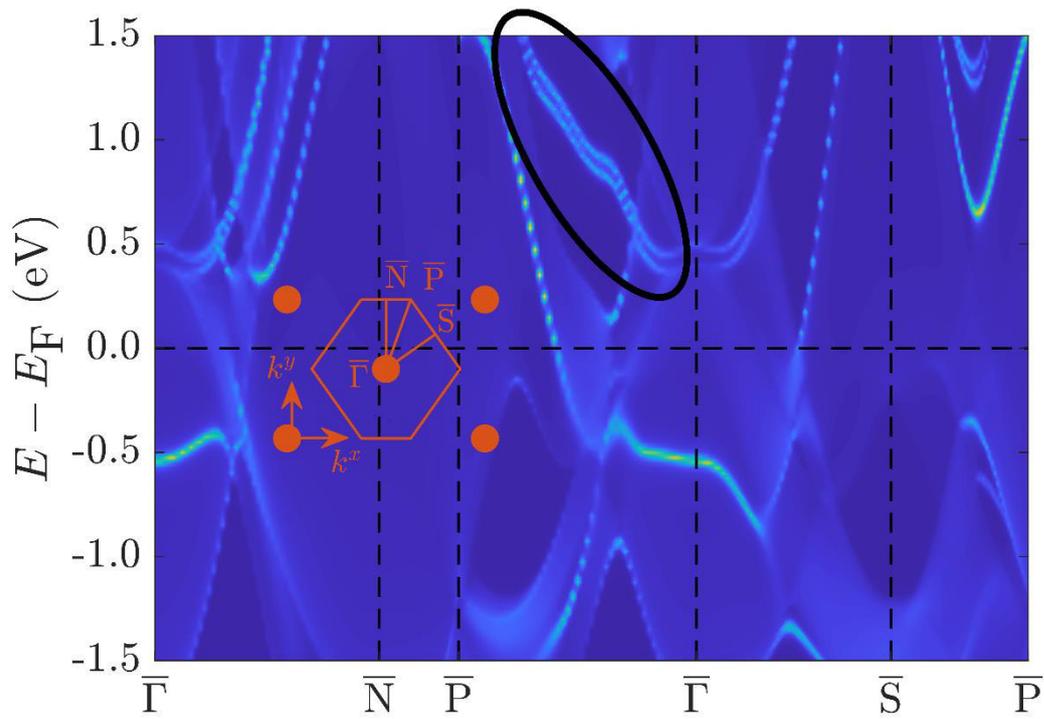

**Supplementary Figure 8 | Surface electronic structure of Nb(110).** Brillouin spectral function from Supplementary Equation (3) calculated along the high-symmetry directions of the Brillouin zone of the Nb(110) surface using the SKKR method. The surface Brillouin zone is illustrated in the sketch in the figure. The black ellipse highlights the surface states used to determine the Rashba parameter. Dashed black lines are guides to the eye.

**Supplementary Note 7. Approximation of the scattering potential strength based on SKKR calculations**

Instead of determining the $K^\mu$ and $J^\mu S/2$ parameters by comparison with the experimental results as described in Supplementary Note 5, they may also be approximated based on the on-site energies and spin splittings of the energy levels of the adatom, along with the strength of the hybridization between the adatom and the substrate, as discussed in refs.[10,16]. These parameters are accessible in first-principles calculations, for example by calculating the density of states at the adatom position.

SKKR calculations based on the embedded cluster algorithm[17] were performed by introducing the Mn adatom in the first vacuum layer of the Nb(110) surface calculated self-consistently in Supplementary Note 6. In the cluster, the 4 Nb atoms were included closest to the Mn adatom in the layer below, as well as 10 empty spheres from the same layer and the layer above. The spin magnetic moment of the adatom was found to be 3.70 μ$_B$, in good agreement with the VASP calculations in Supplementary Note 5. The magnetic easy axis of the adatom was found to lie along the out-of-plane direction, preferred by 0.15 meV over the [001] and by 0.27 meV over the [1$\bar{1}$0] direction, as determined from rotating the magnetic moments in the whole cluster and calculating the band energy based on Lloyd's formula[18]. This supports the out-of-plane magnetic alignment of the adatom and the dimers, assumed in the main text. It was confirmed that these values changed by less than 1% when increasing the cluster size up to 291 atoms.

The DOS on the Mn adatom and in the top Nb layer without the adatom were calculated fixing the imaginary part to 68 meV and changing the energy along the real axis. They are shown in Supplementary Figs. 9a and b, respectively. Spin–orbit coupling was turned off in these calculations, such that the eigenstates may be characterized by spin and angular momentum quantum numbers. Only the projections on the real $l = 2$ spherical harmonics, or $d$ states, are displayed in Supplementary Fig. 9. One or more resonance peaks may be observed on the adatom in each orbital channel, indicating that the atomic orbitals of Mn hybridize with the metallic Nb substrate and split in energy due to the crystal field. All resonance peaks are located below the Fermi level in the majority spin channel and above it in the minority channel, as expected for Mn with a half-filled $d$ band. The hybridization strength depends on the orbitals. For example it is notably weak for the $d_{xy}$ and $d_{x^2-y^2}$ orbitals which overlap less with the Nb atoms in the layer below, leading to sharp resonance peaks. Multiple peaks may be observed for several orbitals, particularly in the majority channel. For the $d_{z^2}$ and the $d_{x^2-y^2}$ orbitals, this may occur because they can hybridize with each other and as such are not eigenchannels, as also observed in the VASP calculations in Supplementary Note 5. For the $d_{yz}$ orbital, this is most likely caused by the hybridization with the substrate which also displays peaks in the DOS in the same energy range (Supplementary Fig. 9b).

The connection between the features in the DOS and the parameters $K^\mu$ and $J^\mu S/2$ is given by the expressions

$$\tilde{J}^\mu = \frac{J^\mu S}{2} \pi \rho_0(E=0) = -\pi \Gamma^\mu \frac{J^\mu}{\left(\varepsilon_d^\mu\right)^2 - (J^\mu)^2} 2S, \qquad \text{Supplementary Equation (5)}$$

$$\tilde{K}^\mu = K^\mu \pi \rho_0(E=0) = -\pi \Gamma^\mu \frac{\varepsilon_d^\mu}{\left(\varepsilon_d^\mu\right)^2 - (J^\mu)^2}, \qquad \text{Supplementary Equation (6)}$$

where $\rho_0(E=0)$ is the DOS in the substrate at the Fermi level, $\pi \Gamma^\mu$ is the half-width at half-maximum of the resonance peaks characterizing the hybridization, and $\varepsilon_d^\mu \pm J^\mu$ are the positions of the peaks in the minority and majority channels, respectively, the splitting of which leads to the emergence of the magnetic moment. Note that analogous expressions may be derived based on the Schrieffer–Wolff transformation[10,16,19],

$$\tilde{\mathcal{J}}^\mu = \pi\Gamma^\mu \frac{U}{\tilde{\varepsilon}_d^\mu(\tilde{\varepsilon}_d^\mu + U)} 2S, \qquad \text{Supplementary Equation (7)}$$

$$\widetilde{K}^\mu = -\pi\Gamma^\mu \frac{2\tilde{\varepsilon}_d^\mu + U}{\tilde{\varepsilon}_d^\mu(\tilde{\varepsilon}_d^\mu + U)}, \qquad \text{Supplementary Equation (8)}$$

where $\tilde{\varepsilon}_d^\mu$ and $\tilde{\varepsilon}_d^\mu + U$ are the energies of the first and the second electrons on the localized *d* level, the latter being increased due to the Coulomb repulsion. Supplementary Eqs. (5) and (6) were deemed more suitable for describing the *ab initio* results, since the exchange–correlation magnetic field in the local spin density approximation of density functional theory breaks the spin symmetry, unlike the interacting Anderson impurity model on which Supplementary Equations (7) and (8) are based. It is important to observe that Supplementary Equations (5) and (6) hold under the assumptions that (i) the scattering channels may be treated together as a single spin of size $S$, (ii) the density of states in the substrate is a constant in the energy range of the order of $\varepsilon_d^\mu$ and $\mathcal{J}^\mu$, and (iii) the shape of the scattering potential is point-like (i.e., $\psi_\mu(\mathbf{k}) \equiv 1$). Under these assumptions, the energies of the Shiba states may be calculated as[20,21]

$$E_{\text{YSR}}^\mu = \pm\Delta \frac{(\tilde{\mathcal{J}}^\mu)^2 - (\widetilde{K}^\mu)^2 - 1}{\sqrt{\left[(\tilde{\mathcal{J}}^\mu)^2 - (\widetilde{K}^\mu)^2 - 1\right]^2 + (2\tilde{\mathcal{J}}^\mu)^2}}. \qquad \text{Supplementary Equation (9)}$$

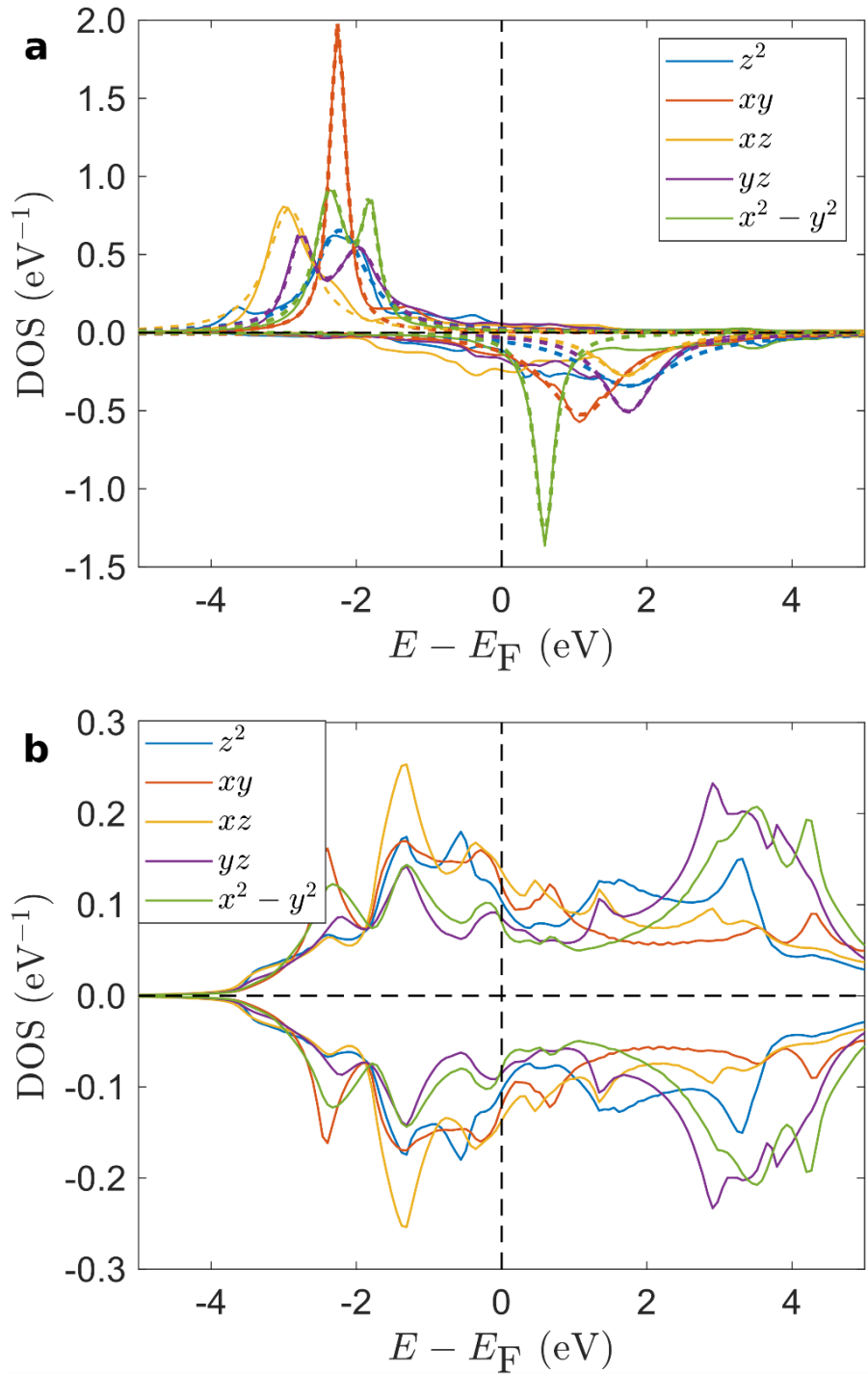

**Supplementary Figure 9 | Density of states in the Mn adatom and the Nb(110) surface layer.** The density of states was calculated from the SKKR method with an imaginary part of the energy of 68 meV. The projection of the DOS on the real *d* orbitals is plotted for **a**, the Mn adatom and for **b**, the top Nb layer (solid curves). The majority and minority spin channels are shown with positive and negative values in the DOS, respectively. Dashed curves in panel **a** denote the Lorentzian curves fitted in the vicinity of the peak positions, used to extract the parameters in Supplementary Equations (5) and (6).

The results obtained from Supplementary Equation (9) are displayed in Supplementary Fig. 10, where the spin size $S$ was used as a scaling parameter. The parameters $\varepsilon_d^\mu$ and $\mathcal{J}^\mu$ were extracted by the peak positions of Lorentzian peaks fitted on the DOS curves shown by dashed lines in Supplementary Fig. 9a. The value of $\pi\Gamma^\mu$ was calculated by determining the width of the Lorentzians, subtracting the artificial broadening of 68 meV introduced in the *ab initio* calculations, and linearly interpolating its value to the Fermi level between the positive and negative energy peaks. From the two peaks of the $d_{x^2-y^2}$ and $d_{yz}$ orbitals in the majority spin channel, the one closer to the Fermi level was selected for the former and the one further away for the latter. It can be deduced from Supplementary Fig. 10, assuming a sufficiently strong coupling such that all Shiba states have crossed the Fermi level, that their energetical order based on the calculations agrees with the experimental one. Note, that the $d_{x^2-y^2}$ state is neglected since only four states were observed experimentally. A quantitative agreement of the Shiba state energies with the experimental values requires assuming $2S = 8 - 10$ for the four selected orbitals. Considering that the assumption of a constant density of states in the substrate in a wide energy range is not supported by the calculations (see Supplementary Fig. 9b) and that the model described by Supplementary Eq. (9) does not take into account the shapes of the scattering potential or of the Fermi surface, we decided to use this description only for a qualitative evaluation of the Shiba states; see refs.[10,22] for similar arguments. Based on these calculations, we concluded that all of the Shiba states are in the limit of strong coupling, which yields the correct order of the Shiba states compared to the experimental situation, and the numerical values of $K^\mu$ and $J^\mu S/2$ were selected to reproduce the Shiba state positions assuming they already crossed the Fermi level.

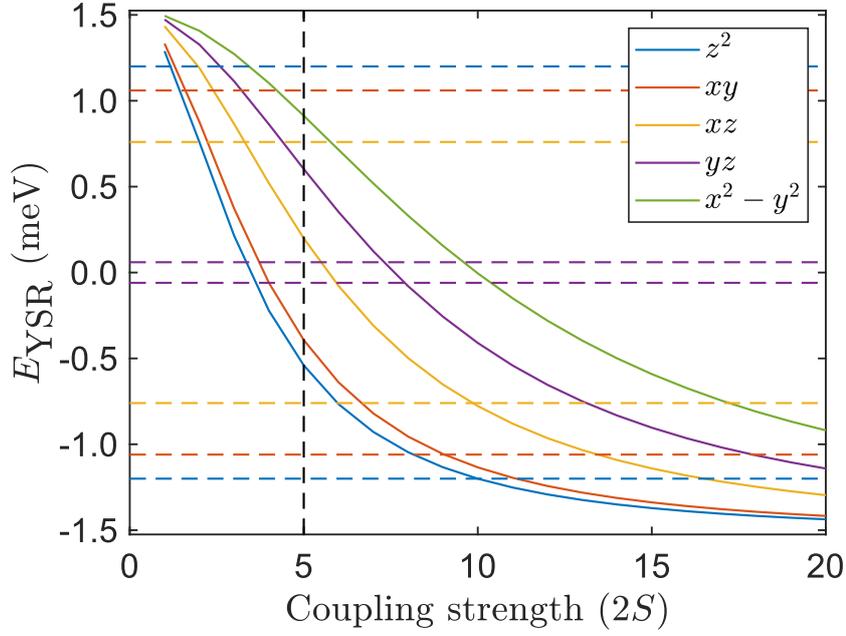

**Supplementary Figure 10 | Energy of Shiba states determined from the SKKR calculations.** Continuous lines denote results from Supplementary Equation (9), with the parameters extracted as described in the text. Horizontal dashed lines denote the experimentally determined positions of the Shiba states, both at positive and negative energy. Shiba states coming from the same orbitals are denoted by the same color in the theoretical curves and in the experimental lines. The corresponding pairs of the Shiba states for the theoretical model are not plotted for simplicity. The total spin of the adatom $S$ was used as a scaling parameter, with the expected value $2S = 5$ for Mn based on Hund's first rule denoted by a black dashed vertical line.

**Supplementary Note 8. Hybridization of Shiba states in the dimers**

Here we discuss the dependence of the splitting of the Shiba states on the angle between the spins and on the strength of the spin–orbit coupling. We consider the $\sqrt{3}a/2 - [1\bar{1}1]$ dimer, and determine the energies of the Shiba states from the condition

$$\det\left[I_4 - \sum_{\mathbf{k}''}\Psi_\mu^*(\mathbf{k}'')G_0(z,\mathbf{k}'')\Psi_\mu(\mathbf{k}'')\frac{1}{N}\left(K^\mu\tau^z - \frac{J^\mu S}{2}S^\alpha\sigma_4^\alpha\tau^z\right)\right] = 0, \qquad \text{Supplementary Equation (10)}$$

cf. Eq. (9) in the Methods section of the main text. Only a single orbital is considered on each adatom in every calculation, thereby neglecting the hybridizations which would occur between different orbitals due to the reduced symmetry. As shown in Supplementary Fig. 11, in the absence of spin–orbit coupling the splitting between the energy positions of the Shiba states monotonically decreases with the angle between the adatom spins, and completely disappears for an antiferromagnetic alignment as expected based on the arguments given in the main text. The only exception to the monotonic decrease can be observed for the δ state, where one of the hybridized orbitals crosses the Fermi level if the angle between the spins is below 140°. Note that the splitting of the β states is significantly smaller than for all the other cases in this specific dimer.

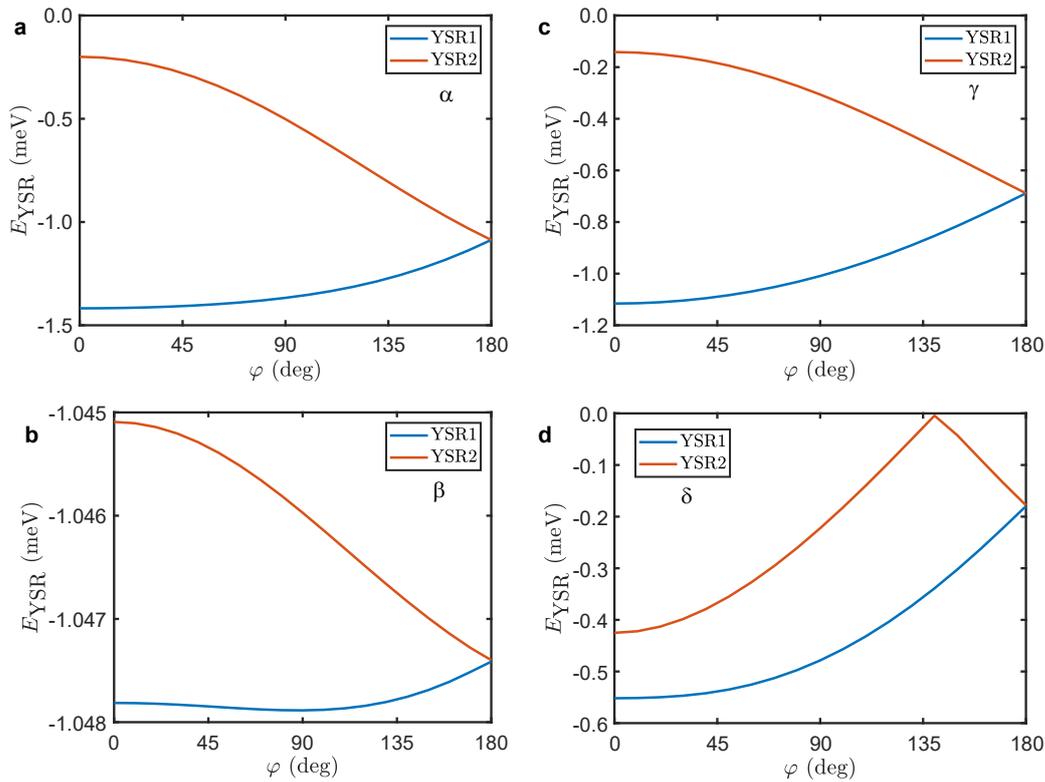

**Supplementary Figure 11 | Hybridization of Shiba states in the $\sqrt{3}a/2 - [1\bar{1}1]$ dimer without spin–orbit coupling.** The scattering parameters are taken from Supplementary Table 2. The orientation of the spins is $\mathbf{S}_1 = (0,0,1)$ and $\mathbf{S}_2 = (\sin\varphi, 0, \cos\varphi)$. The Shiba state positions were calculated separately for **a**, the α, **b**, the β, **c**, the γ and **d**, the δ states.

When spin–orbit coupling is taken into account, the degeneracy of the states in the AFM case is lifted, as shown in Supplementary Fig. 12. Remarkably, the splitting of the β states is considerably larger in the antiferromagnetic alignment than in the ferromagnetic one, and it almost vanishes in the latter case, similar to the result without spin–orbit coupling in Supplementary Fig. 11b. Conversely, the splitting in the ferromagnetic alignment is higher than in the antiferromagnetic one for the other three Shiba orbitals, but even here the angle for which the splitting is maximal differs between the states. This points towards the observation that the effect of spin–orbit coupling is not equivalent to rotating the spin directions on the different sites, as is the case in the one-dimensional Rashba model; see, e.g., ref.[23].

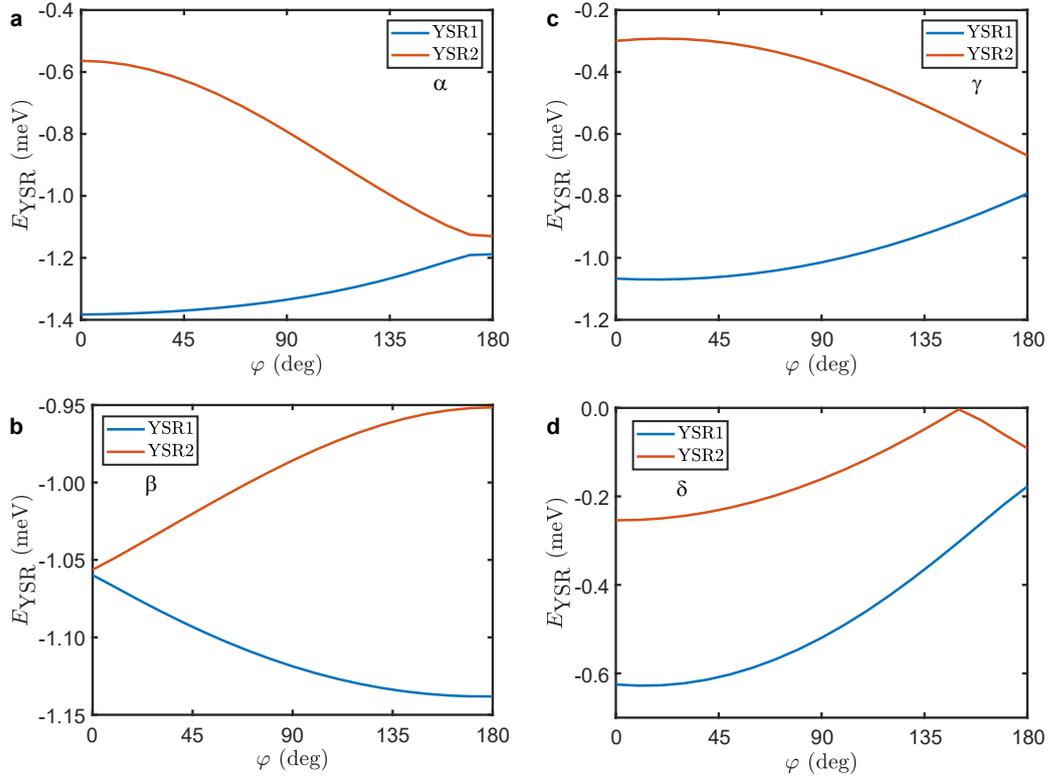

**Supplementary Figure 12 | Hybridization of Shiba states in the $\sqrt{3}a/2 - [1\bar{1}1]$ dimer with spin–orbit coupling.** The scattering parameters are taken from Supplementary Table 1. The orientation of the spins is $\mathbf{S}_1 = (0,0,1)$ and $\mathbf{S}_2 = (\sin\varphi, 0, \cos\varphi)$. The Shiba state positions were calculated separately for **a**, the α, **b**, the β, **c**, the γ and **d**, the δ states.

| dimer | $\varphi_0$ (°) |
|---|---|
| $\sqrt{3}a/2 - [1\bar{1}1]$ | 179.462 |
| $\sqrt{2}\,a - [1\bar{1}0]$ | 179.931 |
| $2\,a - [001]$ | 172.244 |

**Supplementary Table 3 | Ground state angles between the spins in the different dimers from SKKR calculations.**

In order to separate the effects of the non-collinear spin alignment and the Rashba term in the Hamiltonian on the splitting of the Shiba states, both of which are caused by the spin–orbit coupling in a non-centrosymmetric structure, the SKKR calculations discussed in Supplementary Note 7 were extended to dimers. The parameters of a classical spin model were extracted based on the spin-cluster expansion[24,25], and the ground state was found based on this spin model by energy minimization. The ground state angle between the spins is summarized in Supplementary Table 3 for the three dimers discussed in the paper. The calculations predict an almost AFM alignment for all three dimers, which agrees with the experimental results for the $\sqrt{3}a/2 - [1\bar{1}1]$ and the $2a - [001]$ dimers, but disagrees for the $\sqrt{2}a - [1\bar{1}0]$ dimer. The deviation from the collinear alignment is below 10° in all cases and only around 0.5° for the $\sqrt{3}a/2 - [1\bar{1}1]$ dimer, meaning that the Dzyaloshinsky–Moriya interaction causing the non-collinearity is weak in the system. Comparing these small angles to the calculation results in Supplementary Fig. 11, it is clear that the non-collinear alignment itself is insufficient for explaining the experimentally observed magnitude of the splittings of the Shiba states.

It is shown in Supplementary Fig. 13 how the splitting of the Shiba states in the antiferromagnetic alignment depends on the strength of the Rashba parameter $t_R$. In the absence of spin–orbit coupling, the states in the dimer are degenerate, although their energy position is slightly shifted compared to Supplementary Fig. 11 due to the different choice of the scattering parameters. As described in Supplementary Note 5, introducing these different scattering parameters was necessary because the presence of spin–orbit coupling changes the DOS at the Fermi level, which influences the Shiba state positions; see, e.g., Supplementary Equations (5), (6) and (9) for a simple example. This modulation of the DOS is enhanced by the limited resolution in reciprocal space and the energy cut-off restricting the states close to the Fermi level, which may partially explain the considerable oscillations of the Shiba state energies in Supplementary Fig. 13 as the spin–orbit coupling is changed. The numerical accuracy is expected to be improved by considering a finer **k** mesh or a larger energy cut-off. The available data already supports that the effect of the single Rashba coupling parameter is not uniform on the different orbitals, as already discussed above for Supplementary Fig. 12. For example, the $\gamma$ states are almost degenerate for $t_R$=10 meV, while the splitting is almost maximal at this value for the other orbitals.

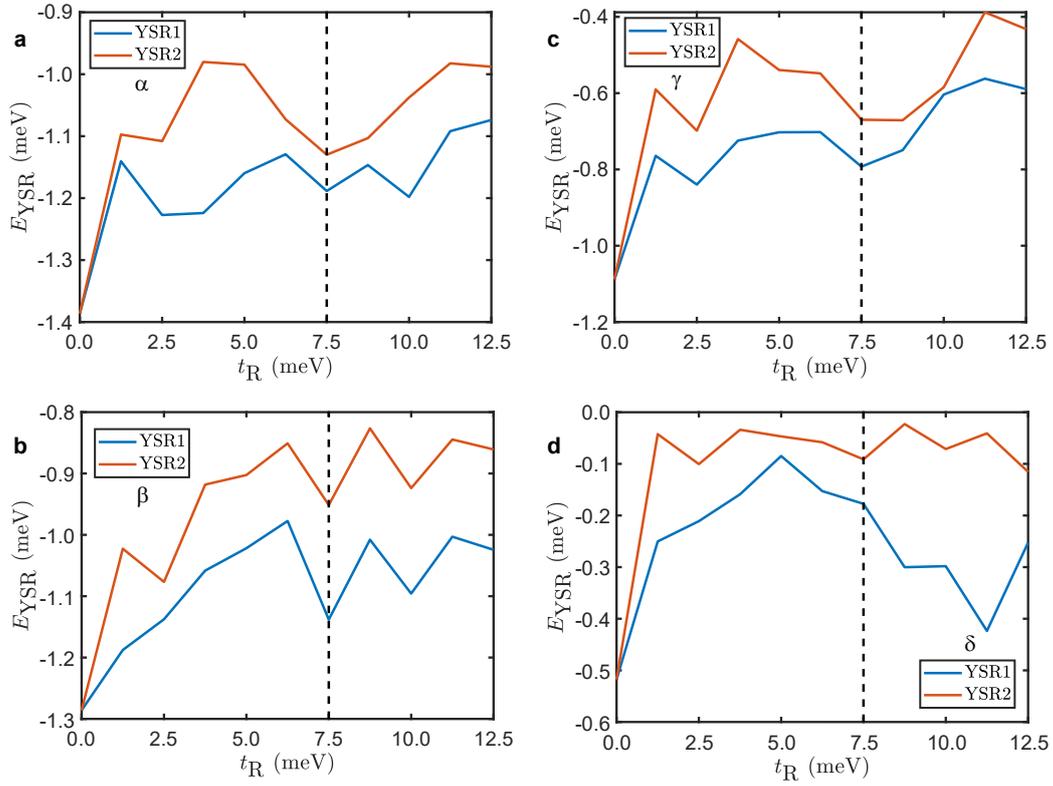

**Supplementary Figure 13 | Hybridization of Shiba states in the $\sqrt{3}a/2 - [1\bar{1}1]$ dimer as a function of spin–orbit coupling.** The scattering parameters are taken from Supplementary Table 1. An out-of-plane antiferromagnetic orientation of the spins was assumed. The Shiba state positions were calculated separately for **a**, the α, **b**, the β, **c**, the γ and **d**, the δ states. Dashed vertical lines denote the value of the Rashba coupling used in Figs. 4f-n in the main text.

## Supplementary Note 9. Symmetry analysis of Shiba state hybridization

The possible hybridizations between Shiba states with different orbitals may be deduced based on group theoretical arguments. The character table of the $C_{2v}$ point group is given in Supplementary Table 4. In this specific case, the character expresses whether the selected *d* orbital is invariant under the selected symmetry operation (1) or it changes sign under it (-1). The possible symmetry operations include the identity transformation $E$, a 180° rotation $C_2$ around the *z* axis, and mirroring $\sigma_v$ on the *xz* and *yz* planes, respectively. The transformations are illustrated in Supplementary Figs. 14a-d for the $d_{xy}$ orbital, which changes sign under both mirror operations but is invariant under the two-fold rotation. There are four possible different ways the orbitals may transform under the action of the group, known as irreducible representations (irreps). Since the Hamiltonian possesses a $C_{2v}$ symmetry, all of its eigenstates, including the Shiba states, belong to one of the irreps, similarly to odd and even eigenfunctions in one dimension. Orbitals belonging to different irreps are always orthogonal and cannot hybridize, but atomic orbitals belonging to the same irrep may hybridize, meaning that the eigenstates of the Hamiltonian will become a linear combination of these atomic orbitals. This can be observed for the $d_{z^2}$ and $d_{x^2-y^2}$ orbitals in the present system, both of which belong to the $A_1$ irrep.

| irrep | $E$ | $C_2$ (z) | $\sigma_v$ (xz) | $\sigma_v$ (yz) | *d* orbitals |
|---|---|---|---|---|---|
| A₁ | 1 | 1 | 1 | 1 | $x^2-y^2$, $z^2$ |
| A₂ | 1 | 1 | -1 | -1 | xy |
| B₁ | 1 | -1 | 1 | -1 | xz |
| B₂ | 1 | -1 | -1 | 1 | yz |

**Supplementary Table 4 | Character table of the $C_{2v}$ point group.**

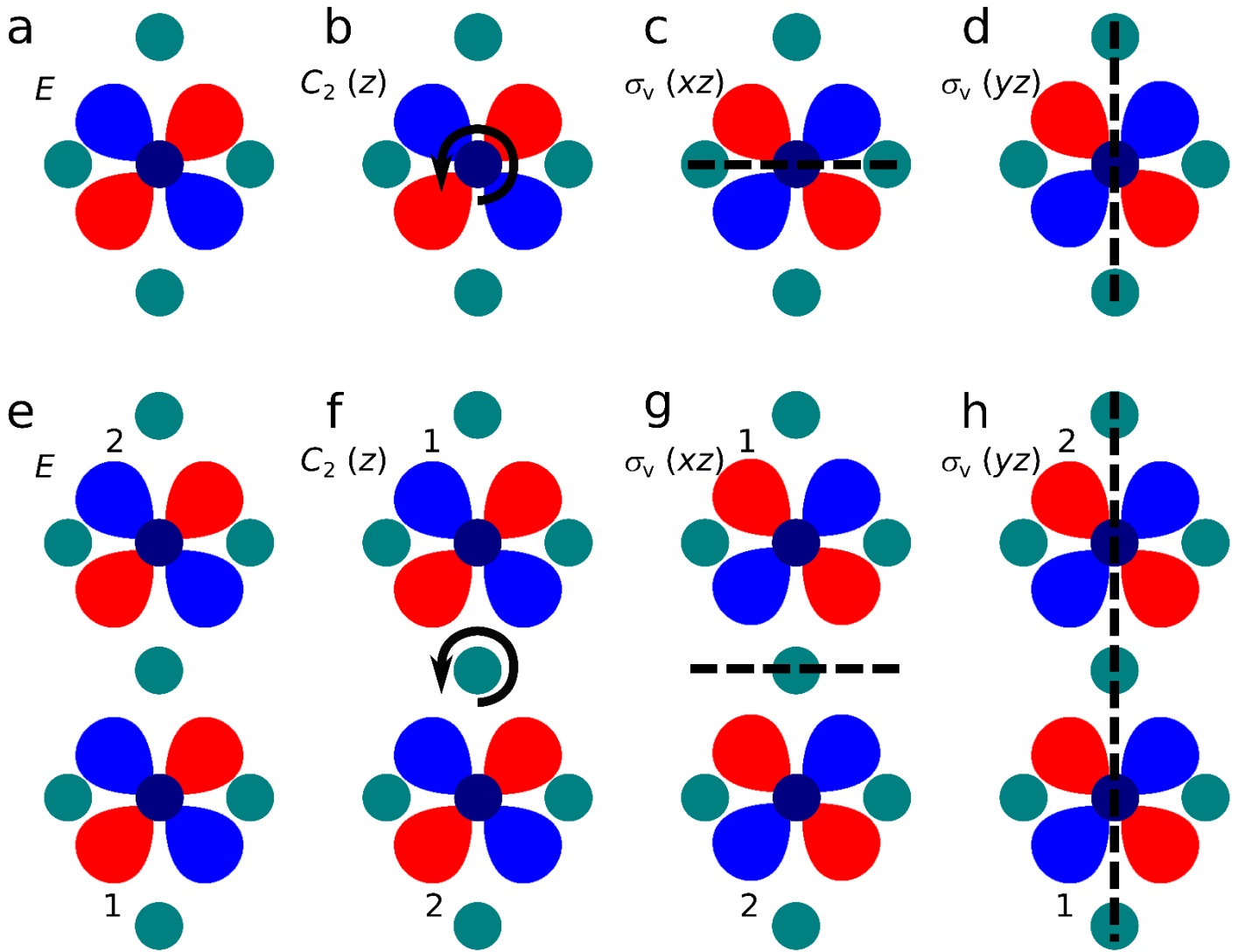

**Supplementary Figure 14 | Transformation of the orbitals under point group operations. a-d** Transformation of an orbital with $d_{xy}$ shape under the symmetry operations included in Supplementary Table 4. Green circles denote atomic positions of Nb atoms. Red and blue lobes indicate positive and negative signs of the wave functions of a Mn adatom, respectively. **e-h** Transformation of two orbitals with $d_{xy}$ shape in a $\sqrt{2}a - [1\bar{1}0]$ dimer arrangement. The numbers 1 and 2 denote the different atomic positions.

The shapes of the Shiba states in the dimers may be deduced by describing them as linear combinations of the adatom orbitals. Under the symmetry transformations given in Supplementary Table 4, both orbitals may stay invariant (1), reverse sign (-1), or move to a different lattice site (0). This is illustrated in Supplementary Figs. 14e-h by considering a $d_{xy}$ orbital on both adatoms in the $\sqrt{2}a - [1\bar{1}0]$ dimer: the wave functions are invariant under $E$ ($\chi = 2$), they change sign under mirroring $\sigma_v$ ($\chi = -2$) on the $yz$ plane, and their positions are exchanged for the twofold rotation and mirroring on the xz plane ($\chi = 0$). Summing up these numbers over the two orbitals give the characters of the selected symmetry transformation, which are listed in Supplementary Table 5. All of these characters, considered as vectors over the different symmetry operations, may be written as linear combinations of the irreps, as given in the last column of Supplementary Table 5. It can be seen that all pairs of orbitals reduce to a pair of bonding ($A_1$, $B_1$) and antibonding ($A_2$, $B_2$) states, which are even and odd under mirroring on the xz plane between the two adatoms, respectively. Eigenstates belonging to the same irreps may still hybridize with each other, meaning that the $d_{z^2}$, $d_{x^2-y^2}$ and $d_{yz}$ orbitals localized at different adatoms will combine, but they will remain orthogonal to the states stemming from the $d_{xy}$ and $d_{xz}$ orbitals.

| adatom orbital | $E$ | $C_2$ (z) | $\sigma_v$ (xz) | $\sigma_v$ (yz) | irreps |
|---|---|---|---|---|---|
| $x^2-y^2$, $z^2$ | 2 | 0 | 0 | 2 | $A_1+B_2$ |
| xy | 2 | 0 | 0 | -2 | $A_2+B_1$ |
| xz | 2 | 0 | 0 | -2 | $A_2+B_1$ |
| yz | 2 | 0 | 0 | 2 | $A_1+B_2$ |

**Supplementary Table 5 | Reduction of atomic orbitals into irreducible representations for the $\sqrt{2}a - [1\bar{1}0]$ dimer.** The same type of orbital is assumed on both adatoms in the dimer.

In the $\sqrt{3}a/2 - [1\bar{1}1]$ dimer, the adatom positions reduce the point group symmetry to $C_2$, with the character table given in Supplementary Table 6. The characters of pairs of atomic orbitals localized at different positions in the dimer have to be calculated according to this symmetry group, and then reduced to irreps, as shown in Supplementary Table 7. In this case, again all pairs of orbitals split into bonding (A) and antibonding (B) states, the latter having a node at the position where the $C_2$ (z) axis crosses the surface. Since all orbitals reduce to the same pair of irreps, any two of them may hybridize for this type of dimer.

| irrep | $E$ | $C_2$ (z) | d orbitals |
|---|---|---|---|
| A | 1 | 1 | xy, $x^2-y^2$, $z^2$ |
| B | 1 | -1 | xz, yz |

**Supplementary Table 6 | Character table of the $C_2$ point group.**

| adatom orbital | $E$ | $C_2$ (z) | irreps |
|---|---|---|---|
| $x^2-y^2$, $z^2$ | 2 | 0 | A+B |
| xy | 2 | 0 | A+B |
| xz | 2 | 0 | A+B |
| yz | 2 | 0 | A+B |

**Supplementary Table 7 | Reduction of atomic orbitals into irreducible representations for the $\sqrt{3}a/2 - [1\bar{1}1]$ dimer.** The same type of orbital is assumed on both adatoms in the dimer.

The above symmetry considerations did not take into account the spin degree of freedom, and they may only be applied if exchanging the atomic positions does not modify the spin configuration (i.e., for ferromagnetic dimers), and in the absence of spin–orbit coupling. Spin–orbit coupling mixes together the different atomic orbitals: for example, a $d_{xy}$ orbital (with magnetic quantum number $m_l = \pm 2$) with spin pointing down ($m_l = -1/2$) is linearly combined with the $d_{xz}$ orbital ($m_l = \pm 1$) with spin pointing up ($m_l = 1/2$) to form an eigenstate of the total angular momentum. Since the orbital part of the up and down spin components transforms differently under the symmetry operations (in the above example, consider the $C_2\,(z)$ rotation in Supplementary Table 6), antibonding states where the total density of states including contributions from both spin channels completely disappears at certain points or along specific lines both for positive and negative energy Shiba states may no longer be strictly defined. This is not immediately apparent for the ferromagnetic dimers if spin–orbit coupling is weak, but it strongly influences the states in antiferromagnetic dimers.